\documentclass[iop,revtex4]{emulateapj}
\newcommand{\myemail}{saito@mpia-hd.mpg.de}
\usepackage[breaklinks,colorlinks,urlcolor=blue,citecolor=blue,linkcolor=blue]{hyperref}

\shorttitle{ALMA Observations of HCN and HCO$^+$ SLEDs toward VV~114}
\shortauthors{Toshiki Saito et al.}

\begin{document}

\title{\bf\normalsize S\lowercase{patially-resolved} D\lowercase{ense} M\lowercase{olecular} G\lowercase{as} E\lowercase{xcitation in the} N\lowercase{earby} LIRG VV~114
}

\author{
   Toshiki \textsc{Saito}\altaffilmark{1,2,3},
   Daisuke \textsc{Iono}\altaffilmark{2,4},
   Daniel \textsc{Espada}\altaffilmark{2,4},
   Kouichiro \textsc{Nakanishi}\altaffilmark{2,4},
   Junko \textsc{Ueda}\altaffilmark{2,5},
   Hajime \textsc{Sugai}\altaffilmark{6},
   Min \textsc{S. Yun}\altaffilmark{7},
   Shuro \textsc{Takano}\altaffilmark{8},
   Masatoshi \textsc{Imanishi}\altaffilmark{9},
   Tomonari Michiyama\altaffilmark{2,4},
   Satoshi \textsc{Ohashi}\altaffilmark{10},
   Minju \textsc{Lee}\altaffilmark{1,2},
   Yoshiaki \textsc{Hagiwara}\altaffilmark{11},
   Kentaro \textsc{Motohara}\altaffilmark{12},
   Takuji Yamashita\altaffilmark{13},
   Misaki Ando\altaffilmark{2,4}
   and
   Ryohei \textsc{Kawabe}\altaffilmark{1,2,4}
}

\email{\myemail}
\altaffiltext{1}{Department of Astronomy, The University of Tokyo, 7-3-1 Hongo, Bunkyo-ku, Tokyo 113-0033, Japan}
\altaffiltext{2}{National Astronomical Observatory of Japan, 2-21-1 Osawa, Mitaka, Tokyo, 181-0015, Japan}
\altaffiltext{3}{Max-Planck Institute for Astronomy, K\"{o}nigstuhl 17, D-69117, Heidelberg, Germany}
\altaffiltext{4}{The Graduate University for Advanced Studies (SOKENDAI), 2-21-1 Osawa, Mitaka, Tokyo 181-0015, Japan}
\altaffiltext{5}{Harvard-Smithsonian Center for Astrophysics, 60 Garden Street, Cambridge, MA 02138, USA}
\altaffiltext{6}{Kavli Institute for the Physics and Mathematics of the Universe (WPI), The University of Tokyo, 5-1-5 Kashiwanoha, Kashiwa, 277-8583, Chiba, Japan}
\altaffiltext{7}{Department of Astronomy, University of Massachusetts, Amherst, MA 01003, USA}
\altaffiltext{8}{Physics Department, College of Engineering, Nihon University, 1 Nakagawara, Tokusada, Tamura, Koriyama, Fukushima, 963-8642, Japan}
\altaffiltext{9}{Subaru Telescope, 650 North A’ohoku Place, Hilo, Hawaii, 96720, USA}
\altaffiltext{10}{RIKEN, 2-1, Hirosawa, Wako-shi, Saitama 351-0198, Japan}
\altaffiltext{11}{Natural Science Laboratory, Toyo University, 5-28-20, Hakusan, Bunkyo-ku, Tokyo 112-8606, Japan}
\altaffiltext{12}{Institute of Astronomy, The University of Tokyo, 2-21-1 Osawa, Mitaka, Tokyo 181-0015, Japan}
\altaffiltext{13}{Research Center for Space and Cosmic Evolution, Ehime University, 2-5 Bunkyo-ku, Matsuyama, Ehime 790-8577, japan}

\begin{abstract}
We present high-resolution observations (0\farcs2--1\farcs5) of multiple dense gas tracers, HCN and HCO$^+$ ($J$ = 1--0, 3--2, and 4--3), HNC~($J$ = 1--0), and CS~($J$ = 7--6) lines, toward the nearby luminous infrared galaxy VV~114 with the Atacama Large Millimeter/submillimeter Array.
All lines are robustly detected at the central gaseous filamentary structure including the eastern nucleus and the Overlap region, the collision interface of the progenitors.
We found that there is no correlation between star formation efficiency and dense gas fraction, indicating that the amount of dense gas does not simply control star formation in VV~114.
We predict the presence of more turbulent and diffuse molecular gas clouds around the Overlap region compared to those at the nuclear region assuming a turbulence-regulated star formation model.
The intracloud turbulence at the Overlap region might be excited by galaxy-merger-induced shocks, which also explains the enhancement of gas-phase CH$_3$OH abundance previously found there.
We also present spatially resolved spectral line energy distributions of HCN and HCO$^+$ for the first time, and derive excitation parameters by assuming optically-thin and local thermodynamic equilibrium (LTE) conditions.
The LTE model revealed that warmer, HCO$^+$-poorer molecular gas medium is dominated around the eastern nucleus, harboring an AGN.
The HCN abundance is remarkably flat ($\sim$3.5 $\times$ 10$^{-9}$) independently of the various environments within the filament of VV~114 (i.e., AGN, star formation, and shock).
\end{abstract}

\keywords{galaxies: individual (VV~114) --- galaxies: interactions --- radio lines: galaxies}

\section{INTRODUCTION}
HCN and HCO$^+$ are known to be excellent unbiased tracers of extragalactic dense molecular interstellar medium \citep[ISM; e.g.,][]{Papadopoulos07,Shirley15}.
Rotational transitions of HCN have high optically-thin critical densities ($n_{\rm crit}$) of $\sim$10$^{5.5}$ cm$^{-3}$, 10$^{7.0}$ cm$^{-3}$, and 10$^{7.4}$ cm$^{-3}$ at the $J$ = 1--0, 3--2, and 4--3 transitions, respectively, when assuming collisional excitation with H$_2$ and a kinetic temperature of 20~K.
Rotational transitions of HCO$^+$, a molecular ion, also have high optically-thin $n_{\rm crit}$ of $\sim$10$^{4.7}$ cm$^{-3}$, 10$^{6.1}$ cm$^{-3}$, and 10$^{6.5}$ cm$^{-3}$ at $J$ = 1--0, 3--2, and 4--3, respectively.
A tight linear correlation in log scale between far-infrared luminosity, $L_{\rm FIR}$ (or star formation rate, SFR), and luminosity of HCN or HCO$^+$ lines, $L'_{\rm dense}$ (or dense gas mass, $M_{\rm dense}$) was found in nearby spiral galaxies and (ultra-)luminous infrared galaxies (U/LIRGs) \citep[e.g.,][]{Gao&Solomon04a,Gao&Solomon04b,Gracia-Carpio08,Garcia-Burillo12,Zhang14,Privon15,Usero15,Stephens16}, suggesting that galaxies have a nearly constant SFR per $M_{\rm dense}$ (i.e., a constant dense gas star formation efficiency; SFE$_{\rm dense}$).
Even if including local star-forming clouds, the relation is still almost linear \citep[e.g.,][]{Shimajiri17}.
Numerical simulations succeeded in reproducing such a linear correlation between log~$L_{\rm FIR}$ and log~$L'_{\rm dense}$ \citep[e.g.,][]{Narayanan08}.
However, \citet{Michiyama16} found a super-linear relation using CO~($J$ = 3--2) line in gas-rich merging galaxies (hereafter CO(3--2) line).

\begin{figure*}
\begin{center}
\includegraphics[width=18cm]{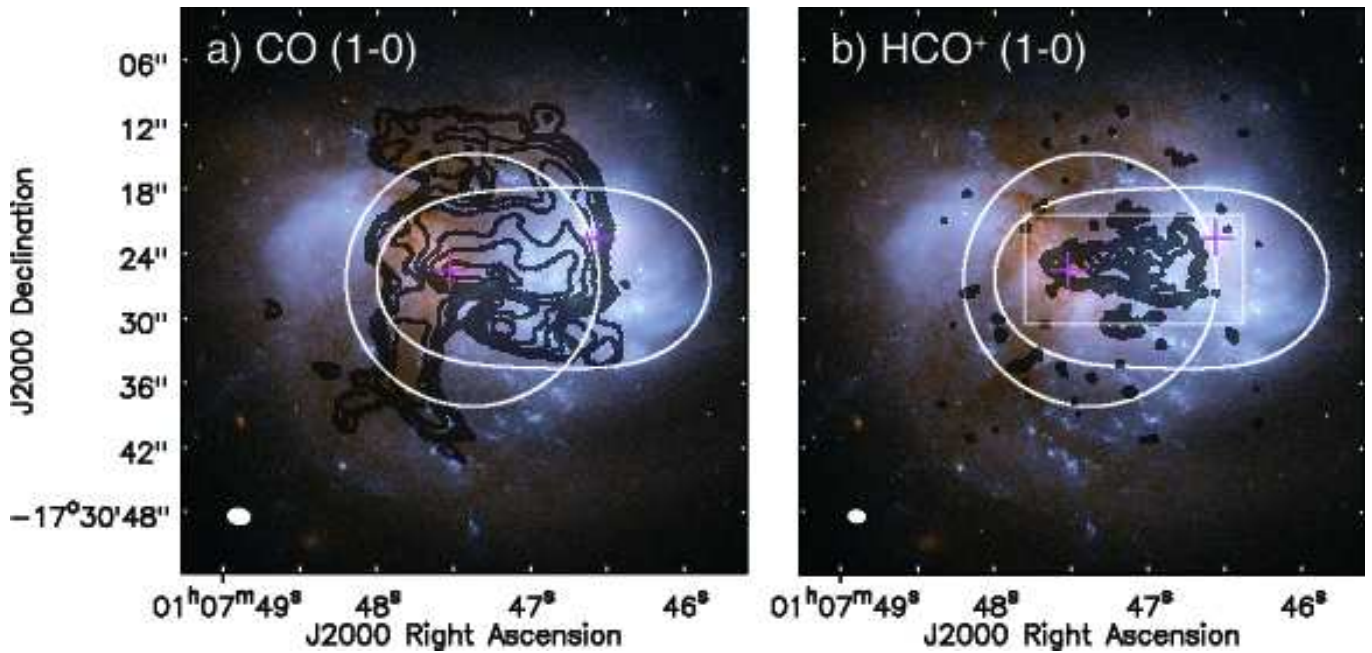}
\caption{(a) ALMA CO~(1--0) contours \citep{Saito15} overlaid on the HST/ACS image of VV~114 [Credit: NASA, ESA, the Hubble Heritage (STScI/AURA)-ESA/Hubble Collaboration, and A. Evans (University of Virginia, Charlottesville/NRAO/Stony Brook University].
The contours are 0.2, 0.4, 0.8, 1.6, 3.2, 6.4, 12.8, 25.6, and 33.0 Jy beam$^{-1}$ km s$^{-1}$.
The white circle and ellipse show the field of view of Band~6 and Band~7, respectively.
The magenta crosses show the positions of the nuclei defined by the peak positions of the 1\farcs0 resolution Ks-band image \citep{Tateuchi15}.
(b) Same as (a), but for HCO$^+$~(1--0).
The contours are 0.67 $\times$ (0.04, 0.08, 0.16, 0.32, 0.64, and 0.96) Jy beam$^{-1}$ km s$^{-1}$.
}
\label{fig_24}
\end{center}
\end{figure*}

\begin{figure}
\begin{center}
\includegraphics[width=8.5cm]{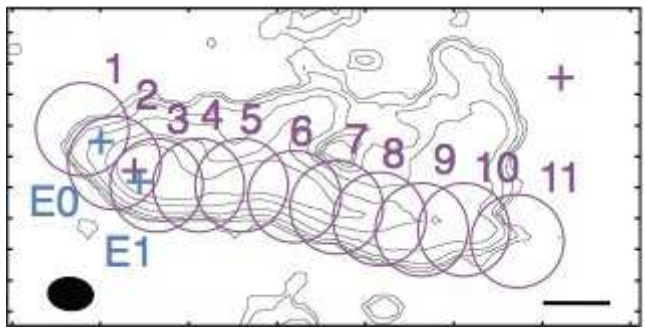}
\caption{The central region of VV~114 as seen in HCO$^+$~(1--0) emission with the aperture and position names used for the flux measurements in this paper;
E0 and E1 are the AGN and starburst (cyan crosses) as defined by \citet{Iono13} and Region 1--11 are apertures along the central dust filament defined by \citet{Saito17b}.
The magenta crosses show the positions of the nuclei defined by the peak positions of the 1\farcs0 resolution Ks-band image \citep{Tateuchi15}.
The scale bar corresponds to 2\farcs0 (= 840~pc).
}
\label{fig_25}
\end{center}
\end{figure}

For multiple CO transitions in U/LIRGs, \citet{Greve14} found a decreasing trend of the $L_{\rm FIR}$-$L'_{\rm CO}$ index as $J$ increases at least up to $J$ = 13--12, which can be explained by a substantial contribution of additional warm and dense gas components to the higher-$J$ line luminosities possibly (mechanically) heated by supernovae, stellar winds and/or active galactic nucleus (AGN) outflow \citep[see also][]{Kamenetzky16}.
On the other hand, the relatively weaker intensities of HCN and HCO$^+$ transitions make it harder in general to study such $J$ dependence for these dense gas tracers for individual galaxies (and from region to region).

Contrary to the analysis of excitation conditions of some molecular species (mostly CO), molecular line intensity ratios between different species have been broadly studied for nearby starburst galaxies and U/LIRGs \citep[][and references therein]{Aalto13}.
Some ratios of bright molecular lines have been proposed to be used to diagnose physical and chemical processes involved in starburst and AGN hidden inside the dusty nuclear regions \citep[e.g., $^{12}$CO/$^{13}$CO, HCN/CO, HCN/HCO$^+$, HCN/HNC, CN/HCN;][]{Aalto91,Kohno01,Aalto02,Gao&Solomon04a,Gao&Solomon04b,Meier05,Costagliola13}.
In particular, the HCN/HCO$^+$ line intensity ratios for a given $J$ ($J_{\rm upp}$ $<$ 5) are among the most established diagnostic tracers of AGN in mm and sub-mm wavelengths, and observational case studies of the ratios have been published for some bright galaxies \citep{Kohno01,Kohno05,Gracia-Carpio06,Imanishi07,Imanishi16a,Imanishi16b,Imanishi16c,Imanishi18,Knudsen07,Papadopoulos07,Krips08,Costagliola11,Hsieh12,Imanishi13a,Imanishi13b,Imanishi14,Iono13,Izumi13,Izumi15,Izumi16,Garcia-Burillo14,Aalto15a,Aladro15,Martin15,Saito15,Schirm16,Ueda16,Espada17,Salak18}.
A few observational and theoretical studies address the excitation state of both molecules \citep[e.g.,][]{Krips08,Meijerink11,Izumi13,Papadopoulos14,Spilker14,Kazandjian15,Tunnard15}.

Understanding the excitation conditions of the dense molecular gas is a natural step to investigate the relation between dense gas, star formation, and AGN.
U/LIRGs are ideal sources to study the excitation of the HCN and HCO$^+$ lines because the high $L_{\rm FIR}$ suggests that dense material is abundant near the nuclear region of these galaxies, and even the high-$J$ transitions of dense gas tracers can be detected with relatively short integration times.
Accordingly, we have used ALMA in the past years to study the dense gas in the merging galaxy VV 114 (\citealt{Iono13,Saito15,Saito16b}, but see also previous studies with other interferometers, \citealt{Yun94,Iono04a,Wilson08,Sliwa13}).

The IR-bright mid-stage merger VV 114 \citep[$D_{\rm L}$ = 87 Mpc, $L_{\rm IR}$ = 10$^{11.69}$ $L_{\odot}$, 1\arcsec = 420~pc;][]{Armus09} is an intriguing system especially for the HCN and HCO$^+$ lines.
Figure~\ref{fig_24}a shows the ALMA CO~(1--0) image ($\sim$1\farcs5 resolution) of VV~114 overlaid on the stellar light taken by {\it HST}/ACS F814W \citep{Saito15}.
The CO emission mainly comes from the central part of VV~114, and the peaks do not coincide with the K-band nuclei (shown as magenta crosses in Figure~\ref{fig_24}).
The dense gas in VV~114 was first imaged in the HCN~(1--0) and HCO$^+$~(1--0) lines with Nobeyama Millimeter Array \citep[5--7\arcsec resolution;][]{Imanishi07}.
The HCN~(1--0)/HCO$^+$~(1--0) flux density ratio is higher than 1.6 at the eastern galaxy.
\citet{Iono13} identified a compact (unresolved by $\sim$0\farcs5 resolution), broad ($\sim$290~km~s$^{-1}$) component in HCN~(4--3) and HCO$^+$~(4--3) lines at the eastern nucleus indicating the presence of a massive object there
($\lesssim$4 $\times$ 10$^8$~$M_{\odot}$).
The unresolved massive object (E0) and the brightest dense gas clump (E1) identified by \citet{Iono13} coincide with the two strongest K-band peaks at the eastern nucleus \citep{Scoville00,Tateuchi15}
This unresolved component was also detected by using Submillimeter Array observations of the HCO$^+$~(4--3) line \citep{Wilson08}.
The HCN~(4--3)/HCO$^+$~(4--3) line ratio at the eastern nucleus was higher than unity, whereas other star-forming dense gas clumps along the central kpc-scale filament of VV~114 showed lower values ($<$ 1).
Combining this information with the observed characteristics in other wavelengths \citep[e.g., Paschen $\alpha$ and hard X-ray;][]{Grimes06,Tateuchi15}, \citet{Iono13} suggested that the eastern nucleus of VV~114 may harbor a dusty AGN (see Figure~\ref{fig_25}), which coincides with a hard X-ray point source \citep{Grimes06}.
Contrary to the eastern nuclei, through our followup CH$_3$OH observations \citep{Saito16b} we found that the Overlap region, which is located between the progenitor's disks of the VV~114 system (i.e., the western side of the kpc-scale filament), is affected by merger-induced shocks.
In summary, VV~114 has three distinctive regions that are expected to show different dense gas excitation and/or chemistry, i.e., the X-ray-bright AGN, starburst regions, and the shocked Overlap, and thus it is a good target to study the properties of dense gas ISM under different environments in a LIRG.
VV~114 is known as the closest analogue of high-z Lyman break galaxies (LBGs) due to its FUV characteristics \citep{Grimes06}.
Since LBGs are thought to constitute a substantial fraction of star-forming galaxies during 2 $\leq$ z $\leq$ 6 \citep{Peacock00}, it is also important to inspect the ISM properties in objects such as VV~114 in order to understand the onset of star formation at high redshift.

This Paper is organized as follows.
Section~\ref{obs} describes a brief summary of our ALMA observations toward VV~114 and the procedure of data reduction.
All HCN and HCO$^+$ lines and continuum images are presented in Section~\ref{results}.
Then, we present beam- and $uv$-matched line ratio images in Section~\ref{ratio}.
 In Section~\ref{discussion}, we discuss the dense gas--FIR luminosity relations and their $J$ dependences (Section~\ref{slope}), star-forming activities using SFEs and dense gas fractions (Section~\ref{SFE}) and its modeling (Section~\ref{SFmodel}), line ratios against SFR (Section~\ref{ratio_discuss}), and the physical properties of dense gas using a radiative transfer model (Section~\ref{LTE}).
 Finally, we summarize our main findings in Section~\ref{conclusion}.
 We have adopted H$_0$ = 70 km s$^{-1}$ Mpc$^{-1}$, $\Omega_{m}$ = 0.3, and $\Omega_{\Lambda}$ = 0.7 throughout this Paper.

\section{OBSERVATIONS AND DATA REDUCTION} \label{obs}
\subsection{Cycle~2 ALMA: J = 1--0} \label{1-0}
The Band~3 line survey toward VV~114 was carried out during the ALMA cycle~2 period (ID: 2013.1.01057.S, PI: T. Saito).
Although the aim of this project was an unbiased line survey (84--111~GHz and 127--154~GHz) to study the chemistry in the filament of VV~114, here we only present two tunings which include the bright dense gas tracers, HCN~(1--0) ($\nu_{\rm obs}$ = 86.88826~GHz), HCO$^+$~(1--0) ($\nu_{\rm obs}$ = 87.43399~GHz), and the ground transition of an isomer of hydrogen cyanide, HNC~(1--0), ($\nu_{\rm obs}$ = 88.88001~GHz) lines.
The full description of this project will be provided in a forthcoming paper.

A tuning (B3--1) covers the HCN~(1--0) and HCO$^+$~(1--0) lines in the lower sideband, and another tuning (B3--2) also covers the HNC~(1--0) line in the lower sideband.
B3-1 (B3--2) was obtained on 2014 June 17, July 2, 2015 June 4, and 5 (2014 July 3) with a single-sideband system temperature ($T_{\rm sys}$) of 38--90~K (33--78~K).
The antenna configuration for B3--1 (B3--2) had thirty to thirty-eight (thirty-one) 12~m antennas, with a projected baseline length ($L_{\rm baseline}$) of 18--780~m (19--650~m), which corresponds to a maximum recoverable scale \citep[MRS;][]{Lundgren13} of $\sim$22\arcsec.
Each tuning had four spectral windows (spws) to cover both sidebands.
Each spw had a bandwidth of 1.875~GHz and 1.938~MHz resolution.
The total on-source time of B3--1 (B3--2) was 47.4~minutes (23.7~minutes).
The field of view (FoV) of Band~3 covers most of the extended CO~(1--0) emission (Figure~\ref{fig_24}a).
Neptune or Uranus were used as flux calibrators, while J0137--2430 or J2258--2758 were used as bandpass calibrators for both tunings.
J0110--0741 or J0116--2052 were observed as phase calibrators.

\subsection{Cycle~3 ALMA: J = 3--2} \label{3-2}
Band~6 observations toward VV~114 were carried out during the ALMA cycle~3 period (ID: 2015.1.00973.S, PI: T. Saito).
The upper sideband was tuned to cover the HCN~(3--2) and HCO$^+$~(3--2) lines.
The data was obtained on 2016 May 23 with $T_{\rm sys}$ of 62--180~K.
The assigned configuration had thirty-seven 12~m antennas with $L_{\rm baseline}$ of 16.7--641.5~m (MRS $\sim$ 9\arcsec).
Each tuning had four spws to cover both sidebands.
Two of the spws including the target lines had a bandwidth of 1.875~GHz with 7.812~MHz resolution, whereas the other two have a bandwidth of 2.000~GHz with 15.625~MHz resolution.
The total on-source time was 30.4 minutes.
The FoV of Band~6 covers all structures found in the HCO$^+$~(1--0) (Figure~\ref{fig_24}b).
Pallas, J0006--0623, J0118--2141 were used as the flux, bandpass, and phase calibrators, respectively.
However, we estimate the absolute flux scaling factor using the bandpass calibrator, because the visibility model of asteroids in {\tt CASA} is not reliable currently (see Section~\ref{reduction}).

\subsection{Cycle~2 ALMA: J = 4--3} \label{4-3}
VV~114 was observed with the long baseline mode of cycle~2 ALMA at Band~7 (ID: 2013.1.00740.S, PI: T. Saito).
The field and frequency setups of the long baseline observations are the same as for our previous HCN~(4--3) and HCO$^+$~(4--3) observations with ALMA \citep{Iono13,Saito15} except for the channel width (976.562~kHz for this Cycle~2 data).
The lower sideband was tuned to cover the CS~(7--6) line.
The data were obtained on 2015 June 28 (B7--1) and July 18 (B7--2) with $T_{\rm sys}$ of 75--205 and 90--310~K, respectively.
The assigned array configuration of B7--1 and B7--2 had fourty-one 12~m antennas with $L_{\rm baseline}$ of 43.3~m--1.6~km and thirty-eight 12~m antennas with $L_{\rm baseline}$ of 15.1~m--1.6~km, respectively.
The combined data has MRS of 7\arcsec.
Each tuning has four spws to cover both sidebands.
All spws have a bandwidth of 1.875~GHz with 976.562~kHz resolution.
The total on-source time of each data is $\sim$41.2 minutes.
J2258-279, J2348-1631, and J0132-1654 were used as the flux, bandpass, and phase calibrators for both data, respectively.

\begin{figure*}
\begin{center}
\includegraphics[width=18cm]{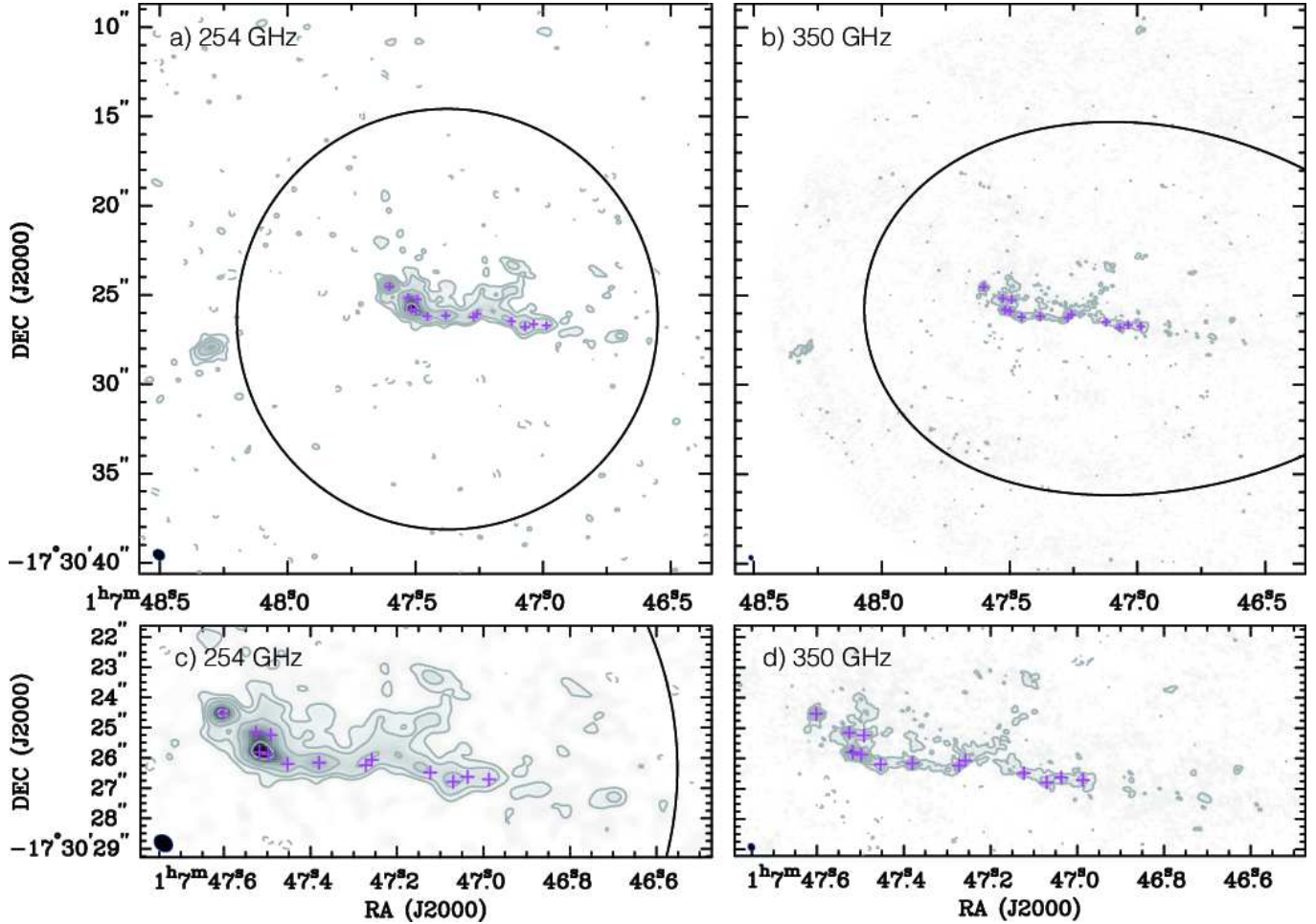}
\caption{(a) The 254~GHz continuum image of VV~114.  The contours are 1$\sigma$ $\times$ (-3, 3, 6, 12, 24, 36, and 48) $\mu$Jy beam$^{-1}$.  1$\sigma$ is 29 $\mu$Jy beam$^{-1}$.  The black line shows the half power beam width (i.e., field of view).  (b) The 350~GHz continuum image of VV~114.  The contours are 1$\sigma$ $\times$ (-3, 3, 6, 12, 24, 36, and 48) $\mu$Jy beam$^{-1}$.  1$\sigma$ is 26 $\mu$Jy beam$^{-1}$.  The black oval shape shows the half power beam width.  Zoomed-in views of (a) and (b) for the filament are shown in (c) and (d), respectively.  All 350~GHz peaks stronger than 8$\sigma$ are marked as crosses.
}
\label{fig_1}
\end{center}
\end{figure*}

\subsection{Data Reduction} \label{reduction}
Processing the new Band~3, Band~6, and Band~7 data, including calibration and imaging, was done using {\tt CASA} version 4.2.2, 4.5.3, and 4.2.2, respectively \citep{McMullin07}.
Images were reconstructed with the natural (robust = 2.0) or briggs (robust = 0.5) weighting.
We made the data cubes with a velocity resolution of 20, 30, or 50 km s$^{-1}$ depending on achieved noise rms and signal-to-noise ratio of the target lines.
Continuum emission was subtracted in the $uv$-plane by fitting the line free channels in both USB and LSB with a first order polynomial function.
The line-free channels were used to make a continuum image using the multi-frequency synthesis method.
Imaging properties, including flux density and recovered flux (i.e., ALMA interferometric flux relative to flux measured by a single-dish telescope), are listed in Table~\ref{table_data}.
We roughly recovered half of the total fluxes measured by single dish telescopes.
All images shown in this Paper, except for line ratios, are not corrected for primary beam attenuation.

We use HCN~(4--3) and HCO$^+$~(4--3) data taken in ALMA cycle~0 \citep[ID: 2011.0.00467.S, PI: D. Iono;][]{Iono13,Saito15}, and combine them with the cycle~2 data (Section~\ref{4-3}).
In those previous papers, we carried out the absolute flux calibrations with the Butler-Horizons-2010 (BH10) models.
However, the flux models were updated, the so-called Butler-JPL-Horizons 2012 (BJH12), so we corrected the previous HCN~(4--3) and HCO$^+$~(4--3) fluxes using the new models following the description in the CASA guides\footnote{https://casaguides.nrao.edu/index.php/Solar$\_$System$\_$Models$\_\\$in$\_$CASA$\_$4.0}.
The flux calibrator used for the cycle~0 data was Uranus, and the difference of the ``zero-spacing" flux density of Uranus between the models (i.e., BJH2012/BH2010 - 1) at 349~GHz is +0.03.
Therefore, we multiplied the flux densities of the cycle~0 data by 1.03 before combination.
Furthermore, we used the {\tt CASA} task {\tt statwt}\footnote{https://casaguides.nrao.edu/index.php/DataWeightsAndComb\\ination} to recalculate the visibility weights of the cycle~0 data.

The systematic error on the absolute flux scaling factor using a solar system object is 5\% for the Band~3 data \citep{Lundgren13}.
As described in Section~\ref{3-2}, we used the bandpass calibrator as the flux calibrator for the Band~6 data in order to avoid using the unreliable flux model of Pallas.
Using the ALMA Calibrator Source Catalogue\footnote{https://almascience.nao.ac.jp/sc/}, the flux uncertainty of J0006-0623 at 260 GHz on 2016 May 23 is estimated to be 7.9\%.
This was estimated by fitting the monitored measurement at 91.5, 103.5 and 343.5~GHz on 2016 May 25 using $S_{\nu}$ $\propto$ $\nu^{\alpha}$, where $S_{\nu}$ is the flux density, $\nu$ is the observed frequency, and $\alpha$ is the spectral index.
The derived $\alpha$ of J0006-0623 is -0.48 $\pm$ 0.01.
Therefore, we adopt a flux uncertainty of the Band~6 data as 8\% throughout this Paper.
Since the flux uncertainties of J2258--279 at 343.5~GHz (i.e., new Band~7 data) on 2015 June 29 and July 20 are $\sim$7.7\% and 15.4\%, respectively, we adopt the flux uncertainty of 15\% for the Band~7 data.

\begin{figure*}
\begin{center}
\includegraphics[width=18cm]{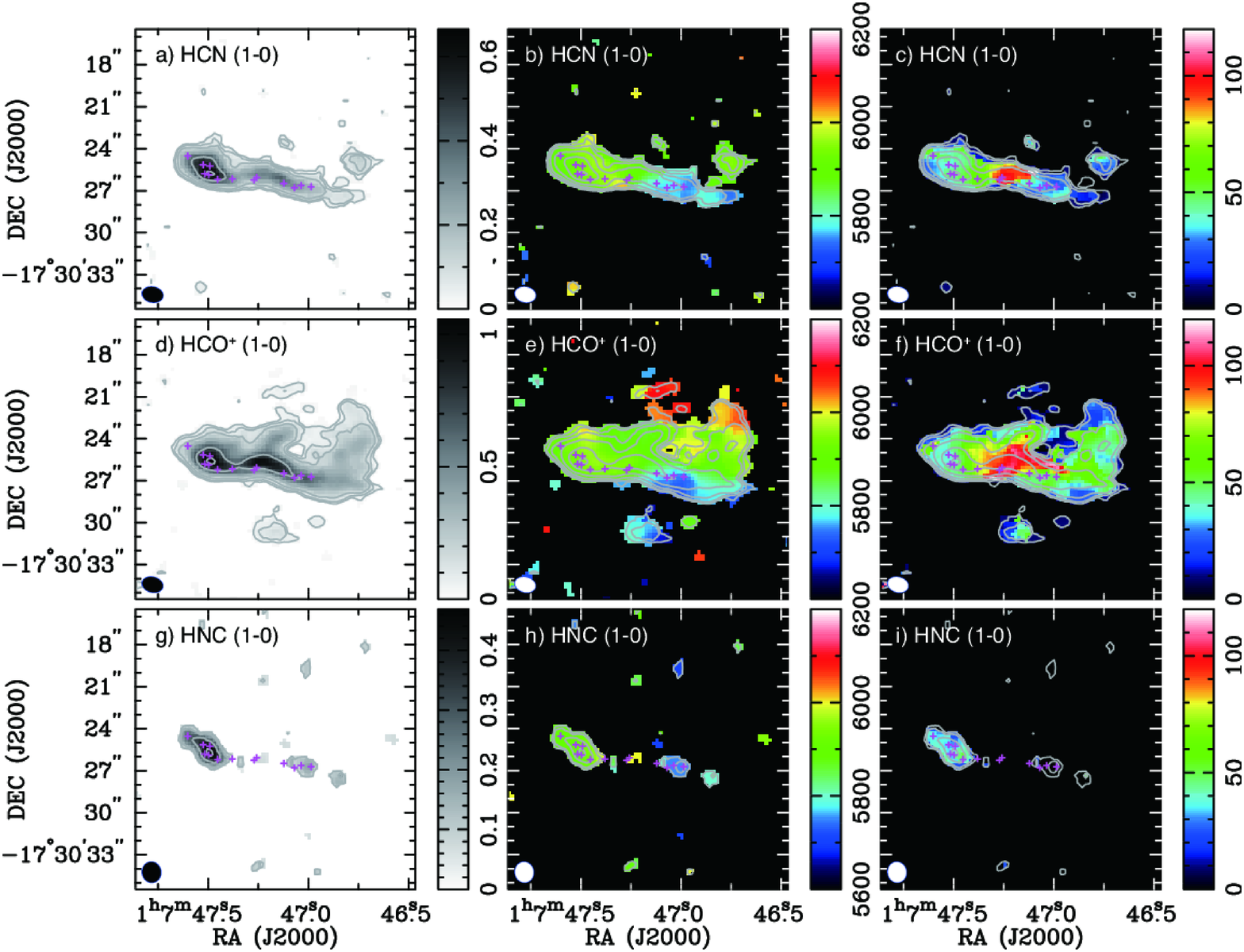}
\caption{(a) Integrated intensity image of HCN~(1--0) of VV~114 in units of Jy beam$^{-1}$ km s$^{-1}$. The synthesized beam is shown in the bottom left corner. The contours are 0.67 $\times$ (0.04, 0.08, 0.16, 0.32, 0.64, and 0.96) Jy beam$^{-1}$ km s$^{-1}$.
The magenta crosses correspond to the 350~GHz peaks stronger than 8$\sigma$ (1$\sigma$ = 26 $\mu$Jy beam$^{-1}$).
(b)  Velocity field image of HCN~(1--0) of VV~114 in units of km s$^{-1}$.  The integrated intensity image is shown as contours.  (c) Velocity dispersion image of HCN~(1--0) of VV~114 in units of km s$^{-1}$.  The integrated intensity image is shown as contours.  (d/e/f) Same as (a/b/c) but for HCO$^+$~(1--0).  The contours are 1.06 $\times$ (0.04, 0.08, 0.16, 0.32, 0.64, and 0.96) Jy beam$^{-1}$ km s$^{-1}$.  (g/h/i)  Same as (a/b/c) but for HNC~(1--0).  The contours are 0.47 $\times$ (0.16, 0.32, 0.64, and 0.96) Jy beam$^{-1}$ km s$^{-1}$.
}
\label{fig_2}
\end{center}
\end{figure*}

If flux measurements for our bandpass and phase calibrators in less than a week from the observing date are available in the catalogue, we check the validity of the absolute flux calibrations using their fluxes for each spw.
For the B3--1 data, the observed fluxes and the monitored values of J0116--2052 and J2258--2758 are consistent ($<$ 9\%).
For the B3--2 data, the flux differences of J0116--2052 are less than 6\%.
The flux differences of J0118--2141, which was used for the Band~6 observation, are less than 30\%.
This is a relatively large value probably because uncertain flux interpolation between Band~3 and Band~7.
The calibrators and check source observed in the Band~7 data were not monitored within 154 days from our observing dates.

MRS of each data is different due to the differences of the observed frequency and the minimum $L_{\rm baseline}$.
When we discuss line ratios between data which have different MRS, we clipped all visibility data inside 18~$k\lambda$ in order to achieve a similar MRS, which allows us to minimize different missing flux effects.
Beam convolution to 1\farcs5 is also performed after $uv$-clipping.

\begin{figure*}
\begin{center}
\includegraphics[width=18cm]{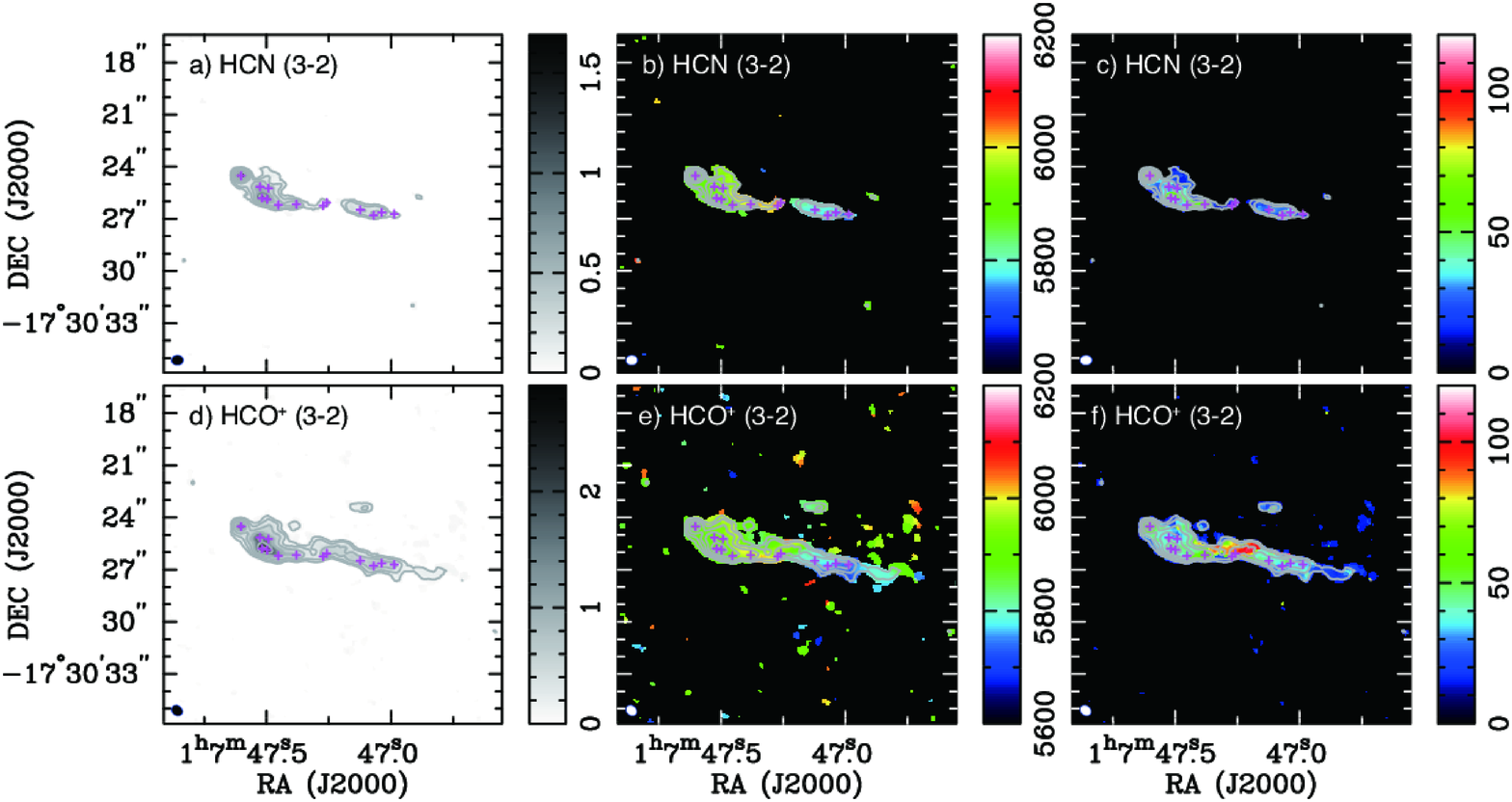}
\caption{(a/b/c) Same as Figure~\ref{fig_2}a, b, and c, but for HCN~(3--2).  The contours are 1.69 $\times$ (0.04, 0.08, 0.16, 0.32, 0.64, and 0.96) Jy beam$^{-1}$ km s$^{-1}$.
The magenta crosses correspond to the 350~GHz peaks stronger than 8$\sigma$ (1$\sigma$ = 26 $\mu$Jy beam$^{-1}$).
(d/e/f) Same as Figure~\ref{fig_2}a, b, and c, but for HCO$^+$~(3--2).  The contours are 2.91 $\times$ (0.04, 0.08, 0.16, 0.32, 0.64, and 0.96) Jy beam$^{-1}$ km s$^{-1}$.
}
\label{fig_3}
\end{center}
\end{figure*}

\begin{figure*}
\begin{center}
\includegraphics[width=18cm]{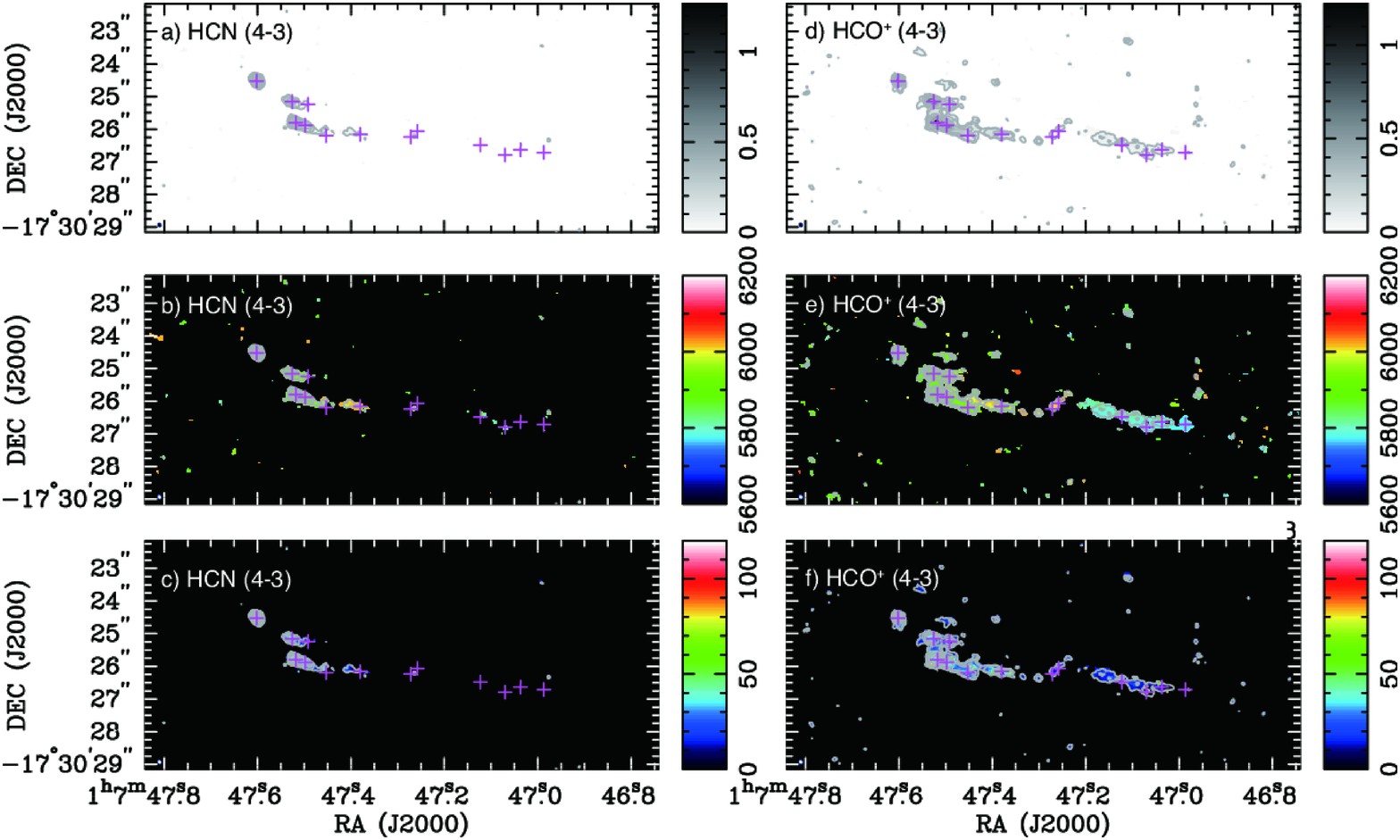}
\caption{(a/b/c) Same as Figure~\ref{fig_2}a, b, and c, but for HCN~(4--3).  The contours are 1.27 $\times$ (0.04, 0.08, 0.16, 0.32, 0.64, and 0.96) Jy beam$^{-1}$ km s$^{-1}$.
The magenta crosses correspond to the 350~GHz peaks stronger than 8$\sigma$ (1$\sigma$ = 26 $\mu$Jy beam$^{-1}$).
(d/e/f) Same as Figure~\ref{fig_2}a, b, and c, but for HCO$^+$~(4--3).  The contours are 1.22 $\times$ (0.04, 0.08, 0.16, 0.32, 0.64, and 0.96) Jy beam$^{-1}$ km s$^{-1}$.
}
\label{fig_8}
\end{center}
\end{figure*}

\section{Results} \label{results}
The 254~GHz (Band~6) and 350~GHz (Band~7) continuum images are shown in Figures~\ref{fig_1}a and \ref{fig_1}b, respectively.
The total flux of VV~114 at 260~GHz and 350~GHz are 17.4 $\pm$ 1.4 mJy and 44.7 $\pm$ 6.7 mJy, respectively.
Their overall spatial distributions along the filament across VV~114 coincide with each other and also with previous 340~GHz (i.e., dust emission mostly) continuum images \citep{Wilson08,Saito15}.
They also agree with the low-resolution 110~GHz (i.e., mostly free-free) \citep{Saito15} and the 1.4~GHz (i.e., synchrotron) radio continuum image \citep{Yun94}.
The filamentary structure consists of a dozen of clumps.
The easternmost point-like source in the filament coincides with the putative AGN \citep{Iono13}.

The integrated intensity, velocity field, and velocity dispersion images of all transitions are shown in Figures~\ref{fig_2}, \ref{fig_3}, and \ref{fig_8}.
All lines are mainly detected at the central filamentary structure which was identified in many molecular lines as well as ionized gas tracers \citep{Tateuchi15}, indicating a site of intense star formation, as already seen in the radio-to-FIR continuum emissions.
The filament consists of three main blobs located at the eastern nucleus, which may harbor a putative AGN (E0, see Figure~\ref{fig_25}) and starbursting clumps (E1), and between the progenitor's disks (``Overlap" region).
For a given $J$, the HCO$^+$ emission is more extended and brighter than the HCN emission.
HNC~(1--0) has two peaks, one at E0 and the another at E1 (see also Section~\ref{E0}).
Although the clumps in the eastern galaxy reported by \citet{Iono13} are clearly resolved in the Band~6 and Band~7 data, they are only marginally resolved in the Band~3 data due to the coarser beam size.
The velocity field and dispersion maps are similar to those of the filament obtained in the CO~(1--0) and $^{13}$CO~(1--0) lines \citep{Saito15}.
The largest velocity dispersion is found around the Overlap region, except for the HNC~(1--0) and HCN~(3--2) images.
The widest velocity dispersion is found around the Overlap region in the HCN~(1--0), HCO$^+$~(1--0), and HCO$^+$~(3--2) images (Figures~\ref{fig_2}c, \ref{fig_2}f, and \ref{fig_3}f).
As clearly seen in the velocity field of HCO$^+$~(3--2) (Figure~\ref{fig_3}f), the wide dispersion ($>$ 100 km s$^{-1}$) is explained by the superposition of the eastern redshifted component and the central blueshifted component (i.e., a double-peaked profile), not due to the intrinsic velocity dispersion of molecular gas clumps there.
We use eleven 3\farcs0 ($\sim$1.2~kpc in diameter) apertures along the filament of VV~114 as in \citet{Saito16b} in order to measure the fluxes of the lines and continuum emission (See Figure~\ref{fig_24}b inset), which are then used for the estimates of molecular gas surface density ($\Sigma_{\rm H_2}$).
The same apertures were adopted to estimate star formation rate surface densities ($\Sigma_{\rm SFR}$) \citep{Saito16b}.
The measured fluxes are listed in Table~\ref{table_flux1}.

\begin{figure*}
\begin{center}
\includegraphics[width=18cm]{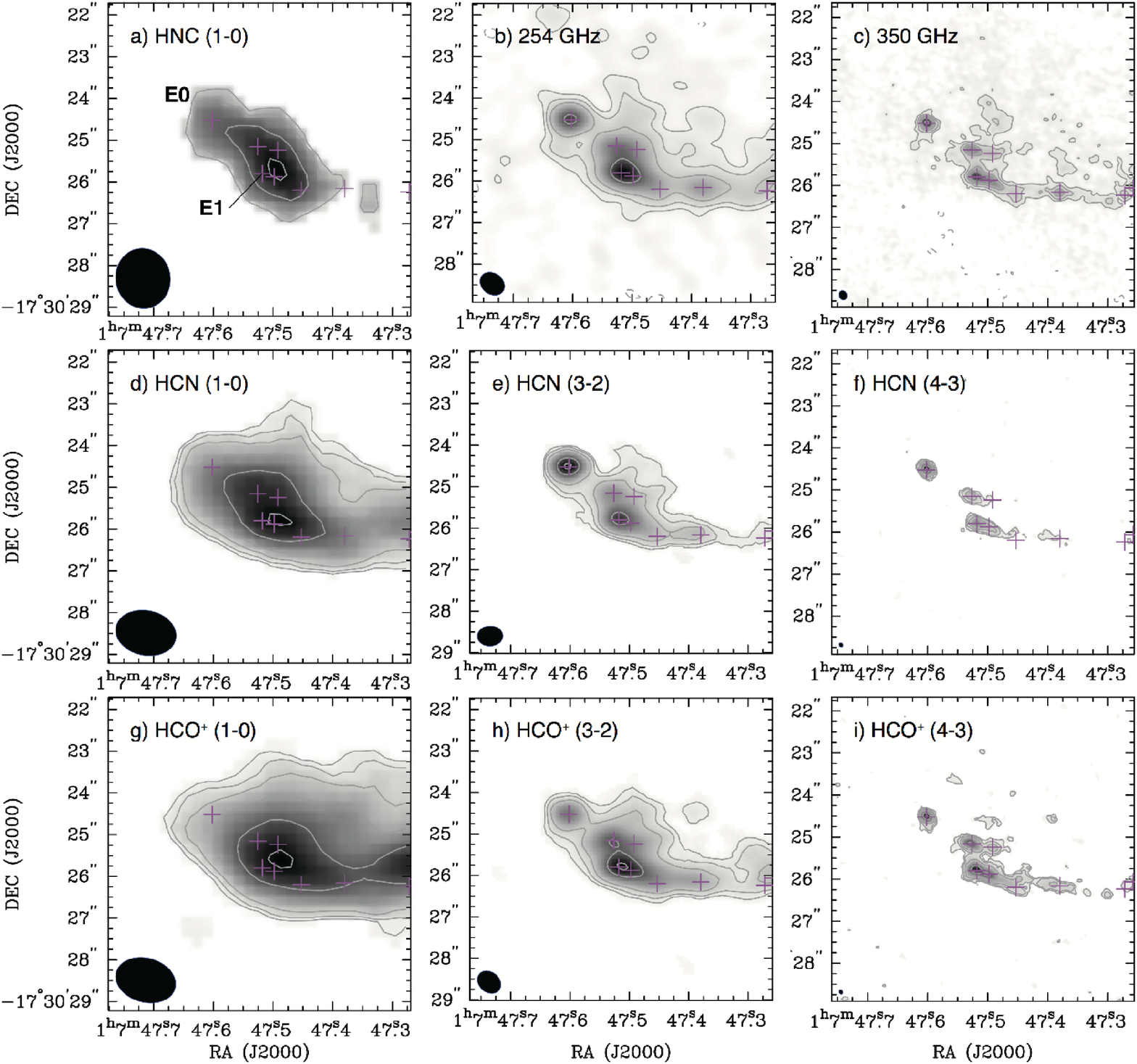}
\caption{Zoomed-in integrated intensity images in the inner 7\arcsec of the eastern galaxy of (a) HNC~(1--0), (b) 254~GHz continuum, (c) 350~GHz continuum, (d) HCN~(1--0), (e) HCN~(3--2), (f) HCN~(4--3), (g) HCO$^+$~(1--0), (h) HCO$^+$~(3--2), and (i) HCO$^+$~(4--3).  The contours are the same as in previous figures.
The magenta crosses correspond to the 350~GHz peaks stronger than 8$\sigma$ (1$\sigma$ = 26 $\mu$Jy beam$^{-1}$).
}
\label{fig_10}
\end{center}
\end{figure*}

\subsection{Clumpy Dense Gas Filament across VV~114}
The 80~pc resolution images of HCN~(4--3) and HCO$^+$~(4--3) emission (Figure~\ref{fig_8}) clearly show clumpy dense gas structures along the filament.
The HCO$^+$ filament has a total length of $\sim$6~kpc and a width of $\lesssim$ 200~pc, and consists of a dozen of giant molecular clouds (GMCs), most of which coincide with the 350~GHz continuum peaks (Figures~\ref{fig_1}b and \ref{fig_1}d).
Such morphological characteristics, together with the enhanced abundance of a molecular gas shock tracer, CH$_3$OH, found at the Overlap region \citep{Saito16b}, share strong similarities with theoretical prediction that a colliding gas-rich galaxy pair forms a filamentary structure at the collision interface (i.e., shock front) between the progenitor's disks \citep[e.g.,][]{Saitoh09,Teyssier10}.

\begin{figure}
\begin{center}
\includegraphics[width=8cm]{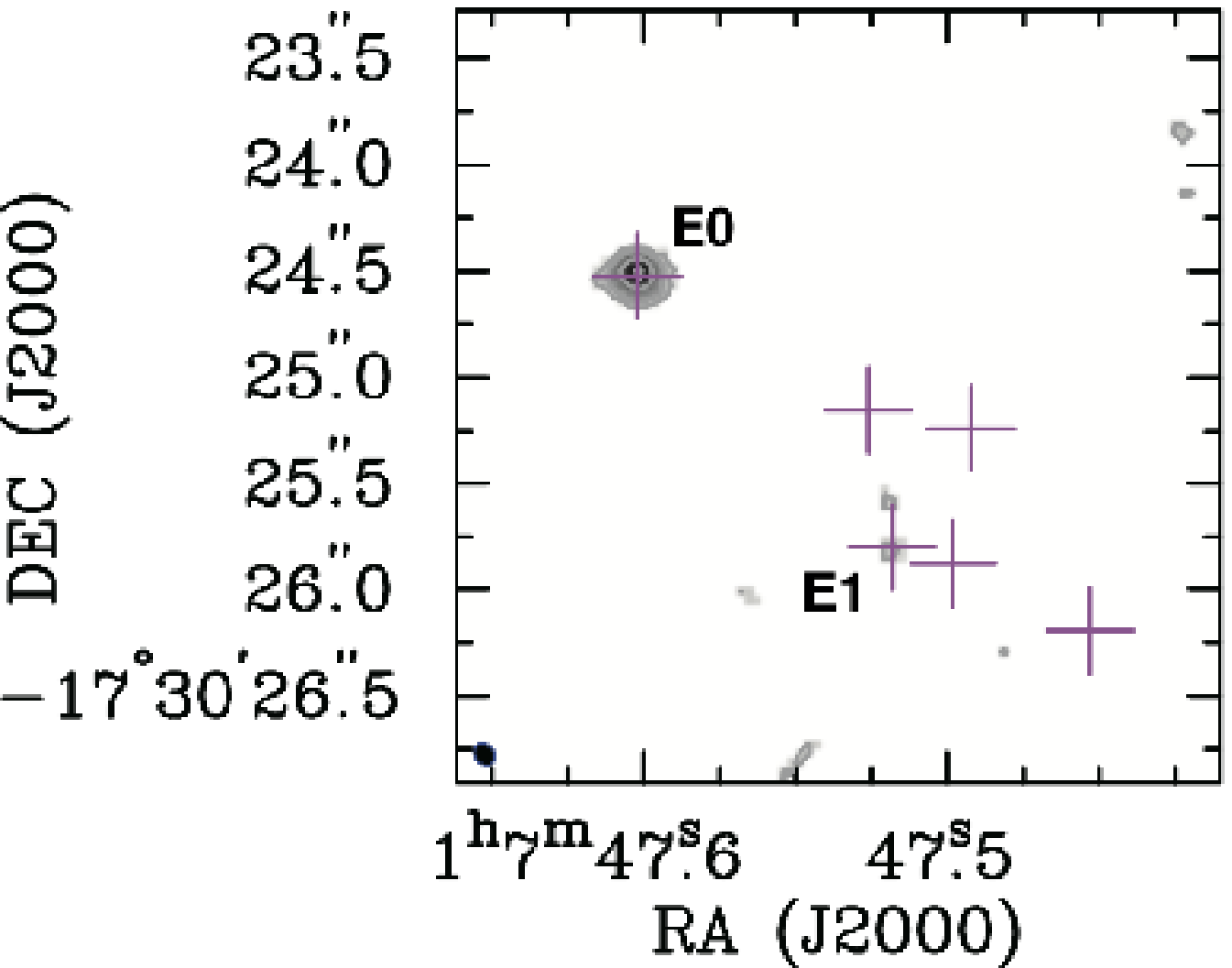}
\caption{Zoomed-in integrated intensity image of CS~(7--6).  The contours are 0.26 $\times$ (0.16, 0.32, 0.64, and 0.96) Jy beam$^{-1}$ km s$^{-1}$.
The magenta crosses correspond to the 350~GHz peaks stronger than 8$\sigma$ (1$\sigma$ = 26 $\mu$Jy beam$^{-1}$).
}
\label{fig_14}
\end{center}
\end{figure}

\begin{figure*}
\begin{center}
\includegraphics[width=16cm]{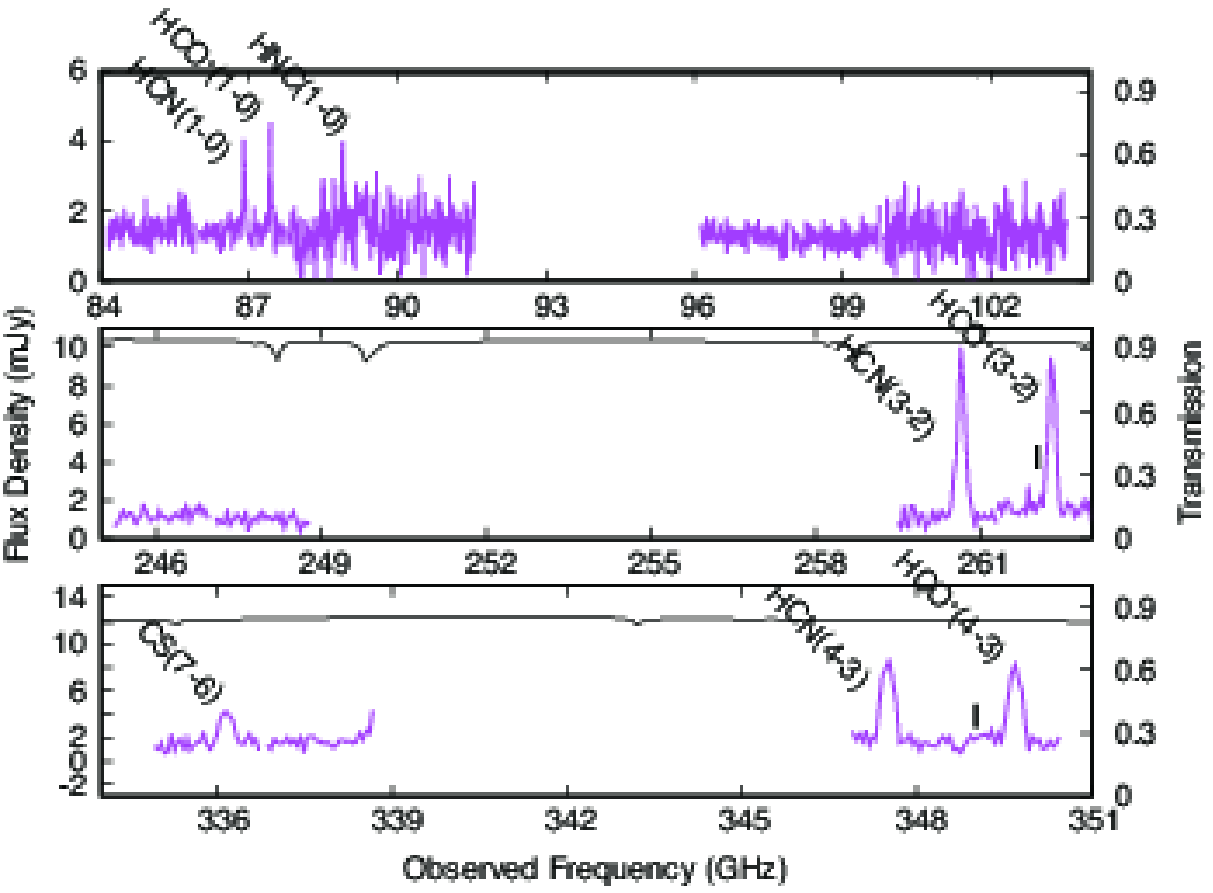}
\caption{Spectra (line + continuum) toward the putative AGN position (E0) at all bands. The velocity resolution is 50 km s$^{-1}$.  Atmospheric transmission with precipitable water vapor of 1~mm is shown in black lines.
The two vertical lines at the left side of HCO$^+$~(3--2) and HCO$^+$~(4--3) show the observed frequency of HCN~($v_2$ = 1$^{1f}$, 3--2) and HCN~($v_2$ = 1$^{1f}$, 4--3), respectively
}
\label{fig_4}
\end{center}
\end{figure*}

\subsection{The Eastern Nucleus} \label{E0}
We show zoom-up images of line and continuum emission around the eastern nucleus in Figure~\ref{fig_10}.
Except for the coarse resolution Band~3 data, the nuclear region is resolved into multiple clumps.
However, most clumps are not resolved even by the 80~pc beam of the Band~7 data.
All line peaks roughly coincide with the 350~GHz continuum peaks, indicating that cold dust grains around the nuclear region are concomitant with dense molecular ISM with 80~pc scale.
In VV~114, the spatial coincidence between dense gas and dust emission is seen from kpc-scale to 80~pc-scale.

Although the Band~7 data provide the highest angular resolution image of molecular line toward VV~114 up to date, the HCN~(4--3), HCO$^+$~(4--3), and 350~GHz continuum images cannot resolve molecular gas structures around the putative AGN, which gives a size upper limit of 80~pc in diameter.
The AGN core also shows bright CS~(7--6) emission (Figure~\ref{fig_14}), which was tentatively detected previously \citep{Saito15}.
Assuming that at the AGN position of all images there is a negligible flux contribution from surrounding star-forming clumps, we measure line and continuum fluxes for the easternmost 350~GHz peak of the filament at ($\alpha$, $\delta$)$_{\rm{J2000}}$ = (01$^{\rm{h}}$07$^{\rm{m}}$47.60, -17\degr30\arcmin24\farcs52), and regard it as molecular ISM which might be dominantly affected by the putative AGN of VV~114.
We measured integrated fluxes at E0 (AGN), as well as E1 (a dense clump at $\sim$1\farcs5 southwest from E0), as defined by \citet{Iono13}, and listed them in Table~\ref{table_flux2}.
Fluxes at E1 might be strongly contaminated by other clumps in the Bands~3 and 6 images, so we only measure line and continuum fluxes in Band~7.
Spectra of all bands toward the AGN position are shown in Figure~\ref{fig_4}.
All lines show broad Gaussian-like profile (FWHM $\sim$ 200~km~s$^{-1}$), indicating a large dynamical mass of $M_{\rm dyn}$ of 9.3 $\times$ 10$^7$~$M_{\odot}$ assuming the inclination of 90\degr$\:$for simplicity.

\begin{figure*}
\begin{center}
\includegraphics[width=18cm]{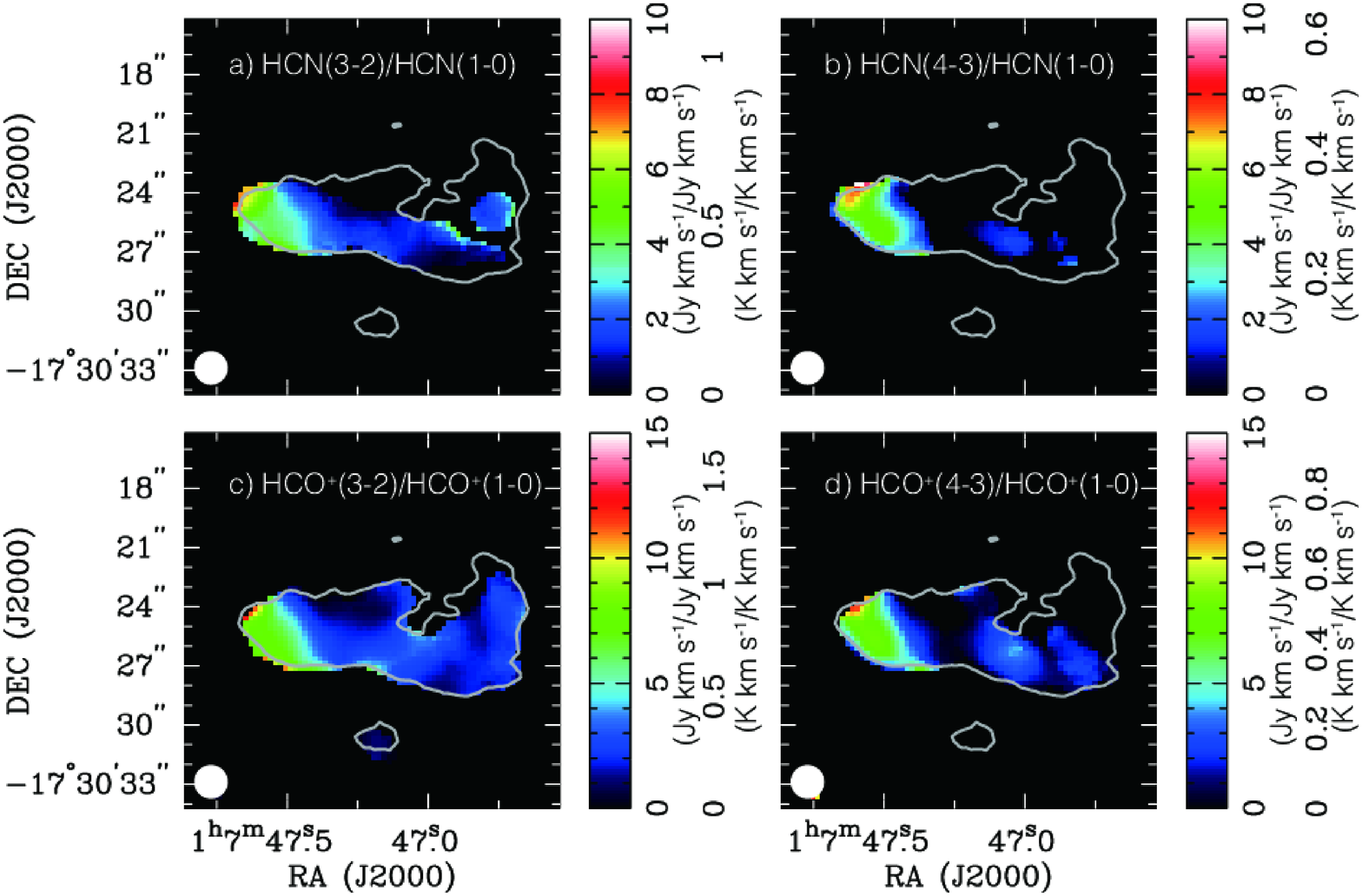}
\caption{(a) HCN~(3--2)/HCN~(1--0) integrated intensity ratio map.  The HCO$^+$~(1--0) contour (= 0.0766~Jy beam$^{-1}$ km s$^{-1}$) is shown as a grey contour.  This contour is the same as the first contour of Figure~\ref{fig_2}d.  (b) HCN~(4--3)/HCN~(1--0) integrated intensity ratio map.  (c) HCO$^+$~(3--2)/HCO$^+$~(1--0) integrated intensity ratio map.  (d) HCO$^+$~(4--3)/HCO$^+$~(1--0) integrated intensity ratio map.  Angular resolution is indicated in the bottom-left of each map.
}
\label{fig_9}
\end{center}
\end{figure*}

\begin{figure}
\begin{center}
\includegraphics[width=9cm]{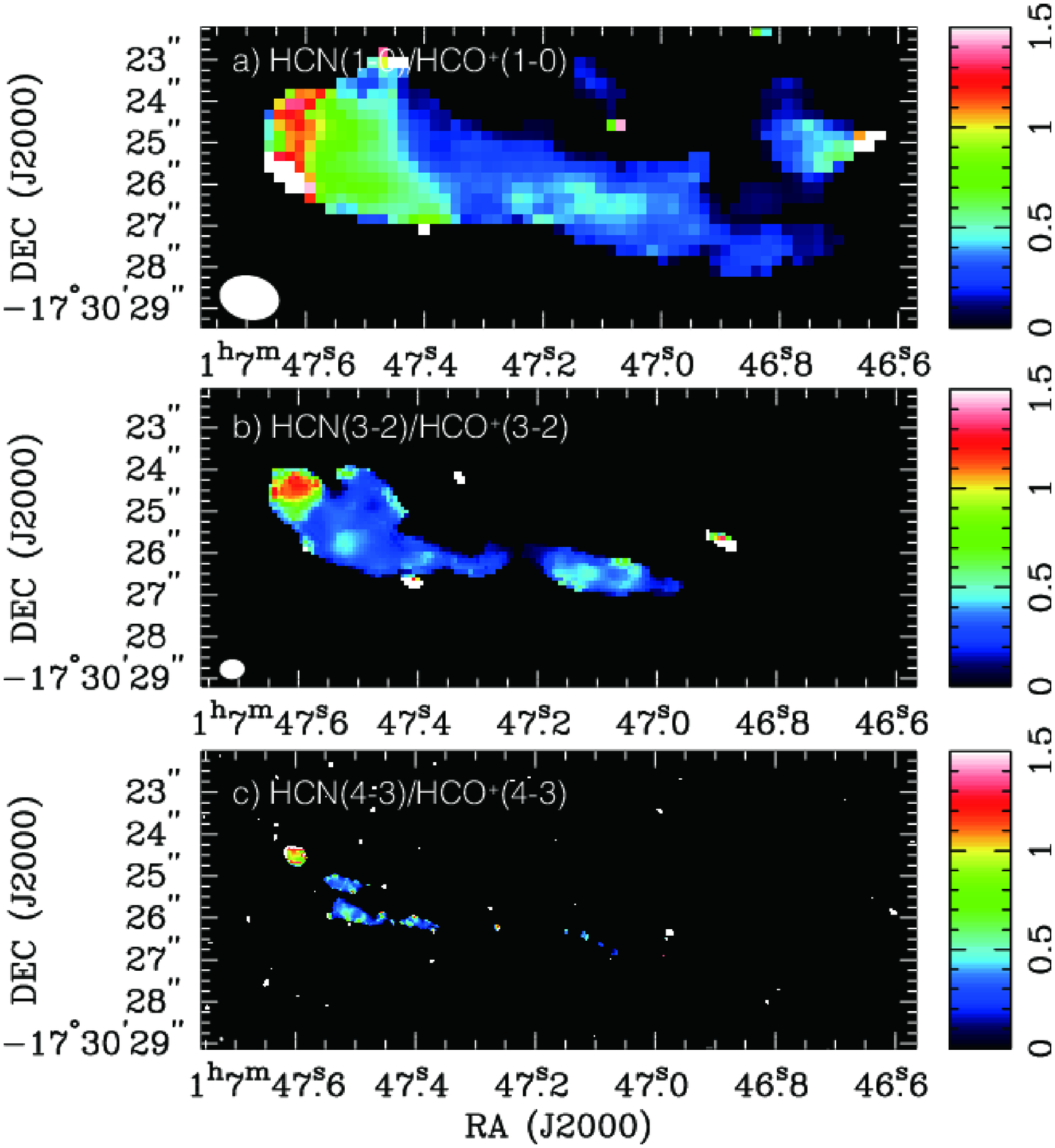}
\caption{(a) Image of HCN~(1--0)/HCO$^+$~(1--0) integrated intensity ratio.  (b) Image of HCN~(3--2)/HCO$^+$~(3--2) integrated intensity ratio.  (c) Image of HCN~(4--3)/HCO$^+$~(4--3) integrated intensity ratio.
}
\label{fig_7}
\end{center}
\end{figure}

\subsection{Vibrationally-excited HCN Lines}
Our frequency coverage contains two vibrationally-excited rotational transitions of HCN, i.e., $J$ = 3--2, $v_2$ = 1$^{1f}$ at $\nu_{\rm rest}$ = 267.1993~GHz ($\nu_{\rm obs}$ = 261.94~GHz) and $J$ = 4--3, $v_2$ = 1$^{1f}$ at $\nu_{\rm rest}$ = 356.2556~GHz ($\nu_{\rm obs}$ = 349.25~GHz).
Those vibrationally-excited HCN lines have been frequently observed toward obscured nuclei in nearby (U)LIRGs \citep[e.g.,][]{Sakamoto10,Sakamoto13,Aalto15b,Imanishi16b,Imanishi17}, and thus they might be detectable toward the putative AGN in VV~114 (E0).
Since the critical densities of those vibrationally-excited HCN lines exceed 10$^{10}$~cm$^{-3}$, it is not easy to excite HCN molecules to a high vibrational states only by collisions.
Instead, radiative excitation, in particular infrared radiation is required to explain detections of the vibrationally-excited HCN lines, and thus those lines are excellent tracers of hot, obscured nuclei in galaxies.

As shown in Figure~\ref{fig_4}, there is no statistically significant feature at these two frequencies.
The 3$\sigma$ upper limit of HCN~($J$ = 3--2, $v_2$ = 1$^{1f}$) and HCN~($J$ = 4--3, $v_2$ = 1$^{1f}$) are 0.26 Jy km s$^{-1}$ and 0.26 Jy km s$^{-1}$, respectively, assuming FWHM = 200 km s$^{-1}$, which is the observed linewidth of HCN~(4--3) and HCN~(3--2) at E0.
The observed flux ratios between $v_2$ = 1$^{1f}$ and $v_2$ = 0 are then $<$0.15 for $J$ = 3--2 and $<$0.16 for $J$ = 4--3.
If compared with the deeply obscured LIRG NGC~4418, which shows 0.17 and 0.23 for $J$ = 3--2 and $J$ = 4--3, respectively \citep{Sakamoto10}, the AGN in VV~114 has weaker $v_2$ = 1$^{1f}$ fluxes normalized by $v_2$ = 0 fluxes.

\section{Line Ratios} \label{ratio}
In this Section, we used the $uv$-clipped and beam-convolved images to construct line ratio images.

\subsection{Excitation Ratios}
The excitation of dense molecular gas in galaxies has been investigated for local galaxies, bright high-z systems, and, in particular, their bright nuclei \citep{Krips08,Izumi13,Papadopoulos14,Spilker14,Tunnard15}, although its spatial variation in LIRGs has not been studied in detail due to the faintness of dense gas tracers.
Here we show spatially-resolved HCN~($J_{\rm upp}$--$J_{\rm upp-1}$)/HCN~(1--0) line ratios and HCO$^+$~($J_{\rm upp}$--$J_{\rm upp-1}$)/HCO$^+$~(1--0) line ratios (hereafter HCN or HCO$^+$ excitation ratios) for a LIRG for the first time (Figure~\ref{fig_9}).
All excitation ratios are made before clipping the inner $uv$-coverage (see Section~\ref{reduction}) and convolving to 1\farcs5 resolution ($\sim$600~pc).
The HCN (HCO$^+$) excitation ratios are masked to the region delimited by the HCN~(1--0) (HCO$^+$~(1--0)) integrated intensity level at 0.095 (0.153) Jy beam$^{-1}$ km s$^{-1}$
because it corresponds to 3$\sigma$ levels in the convolved integrated intensity maps.

The overall spatial distributions of the HCN and HCO$^+$ excitation ratios in VV~114 are coincident with each other.
In all cases, the peak is located at the eastern part of the filament.
All excitation ratios do not exceed the optically-thick thermalized value (unity when using K km s$^{-1}$ units) anywhere.
The highly-excited region at the eastern nucleus extends from northeast to southwest, and coincides with the HNC~(1--0) distribution (Figure~\ref{fig_10}a).
The Overlap region is also relatively excited, but a few times lower than the ratios around the eastern nucleus.
The excitation of diffuse molecular gas phase traced by lower-$J$ CO or $^{13}$CO excitation ratios \citep{Sliwa13,Saito15} shows a similar distribution along the filament with a peak at the eastern edge, indicating that the heating source of the diffuse gas is similar to that of the dense gas.

\subsection{\texorpdfstring{HCN/HCO$^+$ Ratios}{HCN/HCO+ Ratios}}
Contrary to the excitation ratios, intensity ratios between HCN and HCO$^+$ for a given $J$ have been studied to characterize obscured nuclear regions (i.e., AGN or starburst), although HCN/HCO$^+$ ratios at multiple $J$ are not broadly studied, except for some nearby bright galaxies \citep{Krips08,Izumi13}, because of the line weakness and the instrumental limitations.
Using the high quality data of multiple $J$ HCN and HCO$^+$ transitions taken for VV~114, we made HCN/HCO$^+$ ratio images at $J$ = 1--0, 3--2, and 4--3 as shown in Figures~\ref{fig_7}a, \ref{fig_7}b, and \ref{fig_7}c, respectively.
Here we used integrated intensity images without correcting the primary beam attenuation, smoothing the synthesized beam, and $uv$-clipping, because the antenna configurations are virtually identical for HCN and HCO$^+$ at a given $J$ (Table~\ref{table_data}).

\begin{figure*}
\begin{center}
\includegraphics[width=18cm]{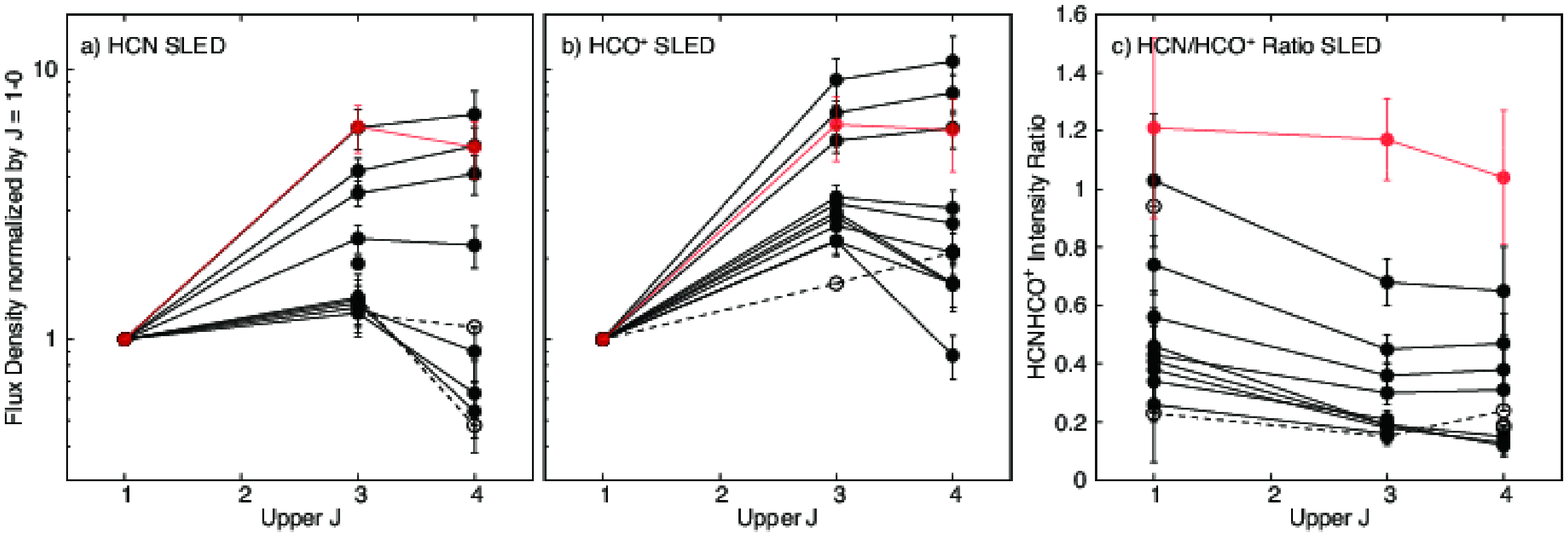}
\caption{The spatially-resolved SLEDs of (a) HCN, (b) HCO$^+$, and (c) HCN/HCO$^+$ ratio.
Black and red lines correspond to data for Region~1-11 and for the AGN position, respectively.
The open circles and dashed lines show 3$\sigma$ upper limits.
Note that the upper three black tracks in all figures correspond to Region~1--3 (i.e., easternmost three apertures).
}
\label{fig_12}
\end{center}
\end{figure*}

All HCN/HCO$^+$ ratio images show the highest peak ($\sim$1.5) at the putative AGN position (E0) despite of the different resolutions.
The second peak is located at $\sim$1\farcs5 southwest of the AGN position, which coincides with a strong peak found in all images shown in Figure~\ref{fig_10} (i.e., star-forming dense clumps).
The second peak value varies from 0.7 to 0.45 as $J$ increases.
The Overlap region shows relatively high values of $\sim$0.4 independently of $J$.
In terms of multiple $J$ HCN and HCO$^+$ line ratios, VV~114, therefore, has three distinctive regions, AGN, star-forming clump(s), and Overlap region, which in turn shows the kpc-scale variation of dense gas excitation and chemistry along the filament.
The higher HCN/HCO$^+$ ratios only near the putative AGN position is consistent with previous studies
\citep[][and references therein]{Garcia-Burillo14,Viti14,Imanishi16c}.
Note that recent high angular resolution observations of HCN and HCO$^+$ lines revealed \citep[e.g.,][]{Martin15,Espada17,Salak18} that the HCN/HCO$^+$ ratios show a variety of spatial variations within circumnuclear disks.

\subsection{HNC/HCN and HCN/CS Ratio}
We detected HNC~(1--0) and CS~(7--6) lines toward the AGN position.
The HNC~(1--0)/HCN~(1--0) line ratio at E0 is 0.85 $\pm$ 0.23, indicating a comparable strength of HNC to HCN at the AGN position.
When using 3\farcs0 apertures (Figure~\ref{fig_24}b), Region 3 shows an HNC~(1--0)/HCN~(1--0) ratio of $\sim$0.25 ($<$ 0.4 for other positions).
The AGN position shows significantly brighter HNC luminosity relative to other molecular regions in VV~114.
HNC/HCN $\gtrsim$ 1 is seen in active nuclei (e.g., \citealt{Huettemeister95,Aalto02,Perez-Beaupuits07,Costagliola11,Aladro15}, see also \citealt{Aalto07} for $J$ = 3--2 transition).
HCN~(4--3)/CS~(7--6) line ratio is suggested to be a diagnostic tracer of AGN activity when combining with the HCN~(4--3)/HCO$^+$~(4--3) line ratio \citep[e.g.][]{Izumi16}.
The HCN~(4--3)/CS~(7--6) ratio is 4.9 $\pm$ 1.3 at the AGN position of VV~114 (E0).
The AGN may have a lower ratio than those in the other regions because it is similar to the lowest 3$\sigma$ lower limit ($>$5.2) of all the other apertures.
On the submillimeter-HCN diagram using HCN~(4--3)/HCO$^+$~(4--3) and HCN~(4--3)/CS~(7--6) ratios \citep{Izumi16}, E0 is similar to NGC~4418 \citep{Sakamoto13} and IRAS12127--1412 \citep{Imanishi14}, both of which may harbor a dust-obscured AGN.

\subsection{SLEDs}
Recent observations have revealed a variety of excitation conditions in dense gas ISM from galaxy to galaxy by comparing HCN and HCO$^+$ spectral line energy distribution (SLED) \citep[e.g.,][]{Rangwala11,Papadopoulos14}.
In Figure~\ref{fig_12}a and \ref{fig_12}b we show spatially-resolved HCN and HCO$^+$ SLEDs, respectively, for the first time in this source, using the eleven apertures (Regions~1-11) and data at the AGN position (E0).
The HCO$^+$ SLEDs of VV~114, in general, tend to show larger values relative to the HCN SLEDs at a given $J$.
Both SLEDs for Region~1-3 and E0 peak at $J$ $\gtrsim$ 3, whereas those for Region~4-11 (i.e., Overlap) peak at $J$ $\lesssim$ 3.
On the other hand, Region~1 (and E0) only show high values ($\gtrsim$ 1) in HCN/HCO$^+$ ratios (Figure~\ref{fig_12}c).
The shape of the HCN/HCO$^+$ ratio with $J$ (hereafter HCN/HCO$^+$ ratio SLED) at E0 is remarkably flat and higher than unity, indicating a specific gas condition relative to other regions.
Since those spatially-resolved SLEDs depend on the variety of thermal and chemical processes in ISM along the filament, we model those in Section~\ref{LTE} to discuss driving mechanisms.

\section{Discussion} \label{discussion}
\subsection{\texorpdfstring{L$_{\rm FIR}$ -- L$'_{\rm dense}$ Relation and Slope -- J Dependence}{LFIR -- L'dense Relation and Slope -- J Dependence}} \label{slope}
Our HCN and HCO$^+$ data have a similar beam size for a given $J$, sensitivity, and maximum recoverable scale, allowing us to probe the dense gas -- star-forming scaling relation without systematic errors to derive fluxes of the dense gas tracers.
In this Paper, we use the SFR surface density ($\Sigma_{\rm SFR}$) based on 110~GHz continuum emission \citep[see][]{Saito16b}, and HCN and HCO$^+$ flux densities as a proxy of $L_{\rm FIR}$ and $L'_{\rm dense}$, respectively.
Here we assumed that free-free (bremsstrahlung) emission dominates the 110~GHz continuum emission, and used the free-free continuum to SFR conversion prescription described in \citet{Yun&Carilli02}.
We note that the derived SFR is upper limit for E0 and Regions 1 and 2, because there might be a contribution from the nonthermal synchrotron emission from the putative AGN.

We roughly estimated the possible synchrotron contribution to the 110~GHz flux at the Overlap region using measured 8.44~GHz and 110~GHz fluxes \citep{Saito15}.
Assuming that the spectral index of the synchrotron emission is -0.80 and that the synchrotron dominates 80\% of the 8.44~GHz flux (similar to M82; \citealt{Condon92}), the synchrotron contribution to the 110 GHz flux is calculated to be $\sim$25\%.
This is not a small value but it does not change any results and discussion in this Paper, because the apertures around the Overlap region also present a similar value of $\sim$25\%.
We need further data points between radio and FIR to precisely evaluate thermal and nonthermal fluxes, which is beyond the scope of this Paper.
We further discuss the slope of $\Sigma_{\rm SFR}$ and flux density relations here.
The ratio (i.e., the star formation efficiency of dense gas, SFE$_{\rm dense}$ = SFR/$M_{\rm dense}$) is discussed in Section~\ref{SFE}.

\begin{figure*}
\begin{center}
\includegraphics[width=18cm]{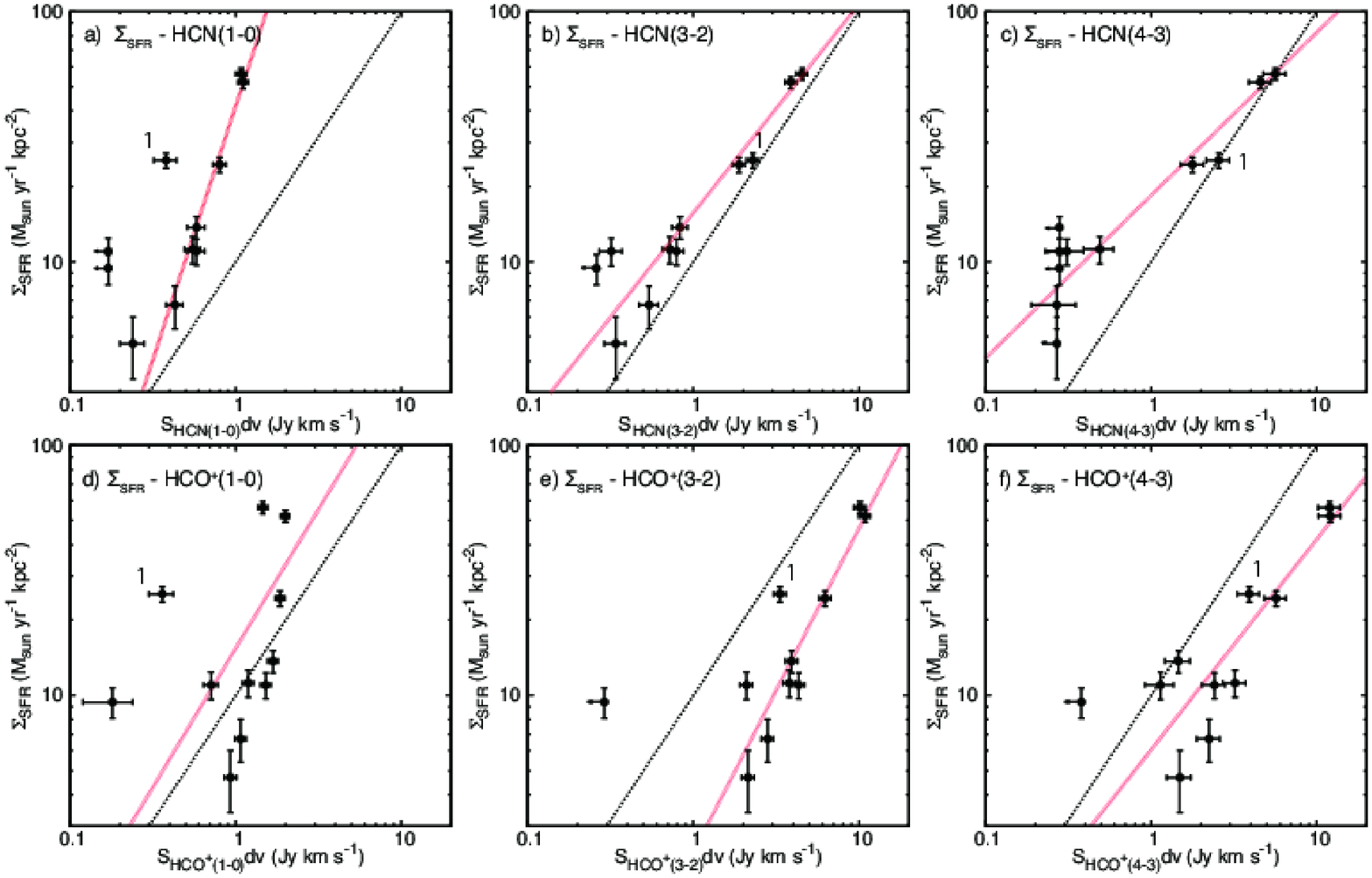}
\caption{Plots of $\Sigma_{\rm SFR}$ against integrated fluxes of all HCN and HCO$^+$ transitions in log scale for the 3\arcsec apertures along the gaseous filament of VV~114 (see the inset of Figure~\ref{fig_24}b).
These are the same as log~$L'_{\rm dense}$-log~$L_{\rm IR}$ plots.
The black dotted line corresponds to a line with a slope of unity.
The data point corresponding to Region 1 is denoted as 1.
The best fitted lines excluding Region 1 are shown as red dotted lines.
}
\label{fig_11}
\end{center}
\end{figure*}

\begin{figure}
\begin{center}
\includegraphics[width=8cm]{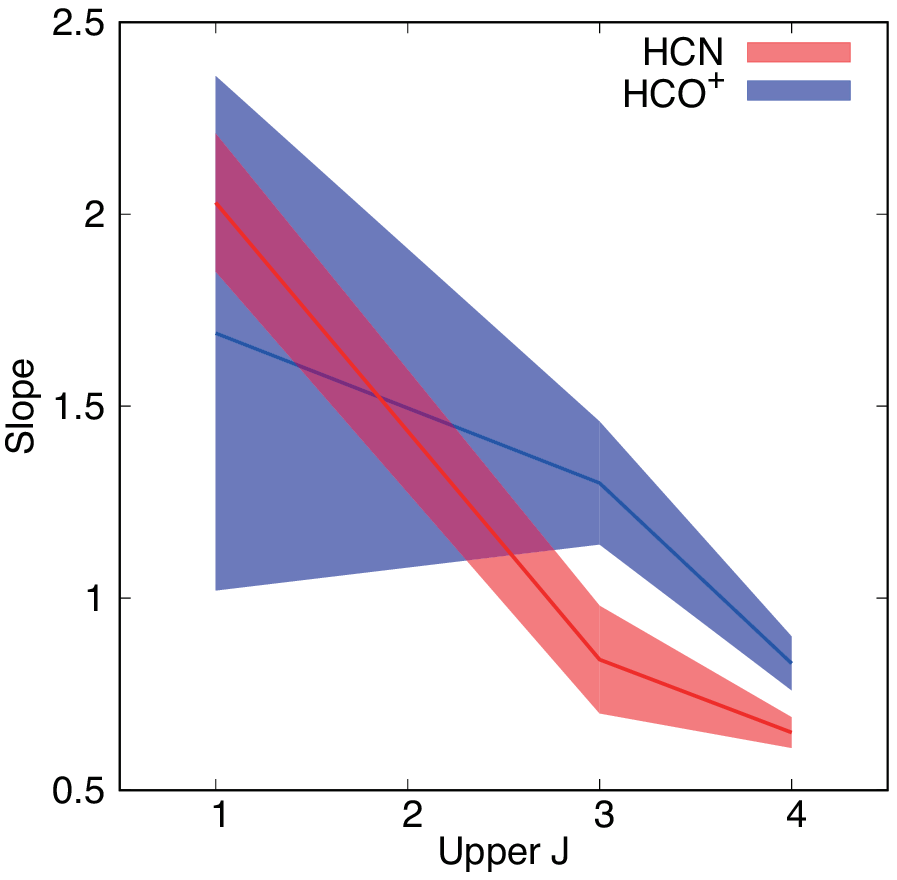}
\caption{The index of the log-log plots shown in Figure~\ref{fig_11}.  Red and blue tracks correspond to HCN and HCO$^+$ indices, respectively.  1$\sigma$ error ranges of the fitting are shown as colorized areas.
}
\label{fig_22}
\end{center}
\end{figure}

In Figure~\ref{fig_11}, $\Sigma_{\rm SFR}$ is plotted against HCN and HCO$^+$ integrated fluxes.
In general, integrated flux densities roughly correlate with $\Sigma_{\rm SFR}$ in log scale as pointed out in many studies \citep[e.g.,][]{Gao&Solomon04a,Gao&Solomon04b,Privon15,Usero15,Shimajiri17}.
We estimate the slope by fitting all the data points except Region~1, where considerable contamination by the putative AGN might be possible.
The derived slope for $\Sigma_{\rm SFR}$ vs. HCN~(1--0), HCN~(3--2), HCN~(4--3), HCO$^+$~(1--0), HCO$^+$~(3--2), and HCO$^+$~(4--3) fluxes are 2.03 $\pm$ 0.18, 0.83 $\pm$ 0.07, 0.65 $\pm$ 0.04, 1.69 $\pm$ 0.67, 1.30 $\pm$ 0.16, and 0.84 $\pm$ 0.14, respectively.
Although both the number of data points and the covered x- and y-ranges are small, the slope vary for different dense gas tracers.
To clarify the excitation dependence, we plot the slope as a function of upper-$J$ in Figure~\ref{fig_22}, showing a decreasing trend as upper-$J$ increases of both HCN and HCO$^+$ (from $\sim$2 to $\sim$0.5).
This can be explained by two simple mechanisms, as we discuss next:
\begin{enumerate}
\item At the lower $\Sigma_{\rm SFR}$ regime ($\lesssim$ 20 $M_{\odot}$ yr$^{-1}$ kpc$^{-2}$), FUV heating by star-forming activity is inefficient and not enough to thermalize high critical density tracers toward higher-$J$.
Consequently, flux densities of $J$ = 3--2 and 4--3 lines of HCN and HCO$^+$ decrease, and thus the index becomes sublinear.
Hydrodynamic simulations with non-LTE radiative transfer calculations \citep{Narayanan08} reproduced this trend for HCN transitions from $J$ = 1--0 to 5--4.
An analytic model provided by \citet{Krumholz&Thompson07} also reproduced this similar trend as a function of critical density.
\item At the higher $\Sigma_{\rm SFR}$ regime ($\gtrsim$ 20 $M_{\odot}$ yr$^{-1}$ kpc$^{-2}$), high critical density tracers can be easily excited by intense FUV radiation from massive star-forming regions (i.e., efficient collisional excitation due to high temperature or density).
Furthermore, the presence of energetic activities (e.g., shock, AGN) tends to excite molecular gas ISM.
Recent multiple $J$ CO observations have revealed that higher $J$ CO lines are more dominantly heated by such energetic activities \citep{Greve14}, and brighter galaxies at IR show more excited CO conditions \citep{Kamenetzky16}.
Expanding those evidences to dense gas tracers, excitation of dense gas ISM at higher $\Sigma_{\rm SFR}$ regime (i.e., higher IR) should be enhanced by energetic activities.
Thus, the presence of intense starbursts and energetic activities can boost fluxes of higher $J$ HCN and HCO$^+$ lines, leading to the sublinear slope.
This is consistent with the explanation of a decreasing index with increasing $J$ of CO due to an increasing contribution of mechanical heating from supernovae \citep{Greve14}.
We will quantitatively discuss the excitation mechanisms of the dense gas tracers in Section~\ref{LTE}.
\end{enumerate}

\begin{figure*}
\begin{center}
\includegraphics[width=18cm]{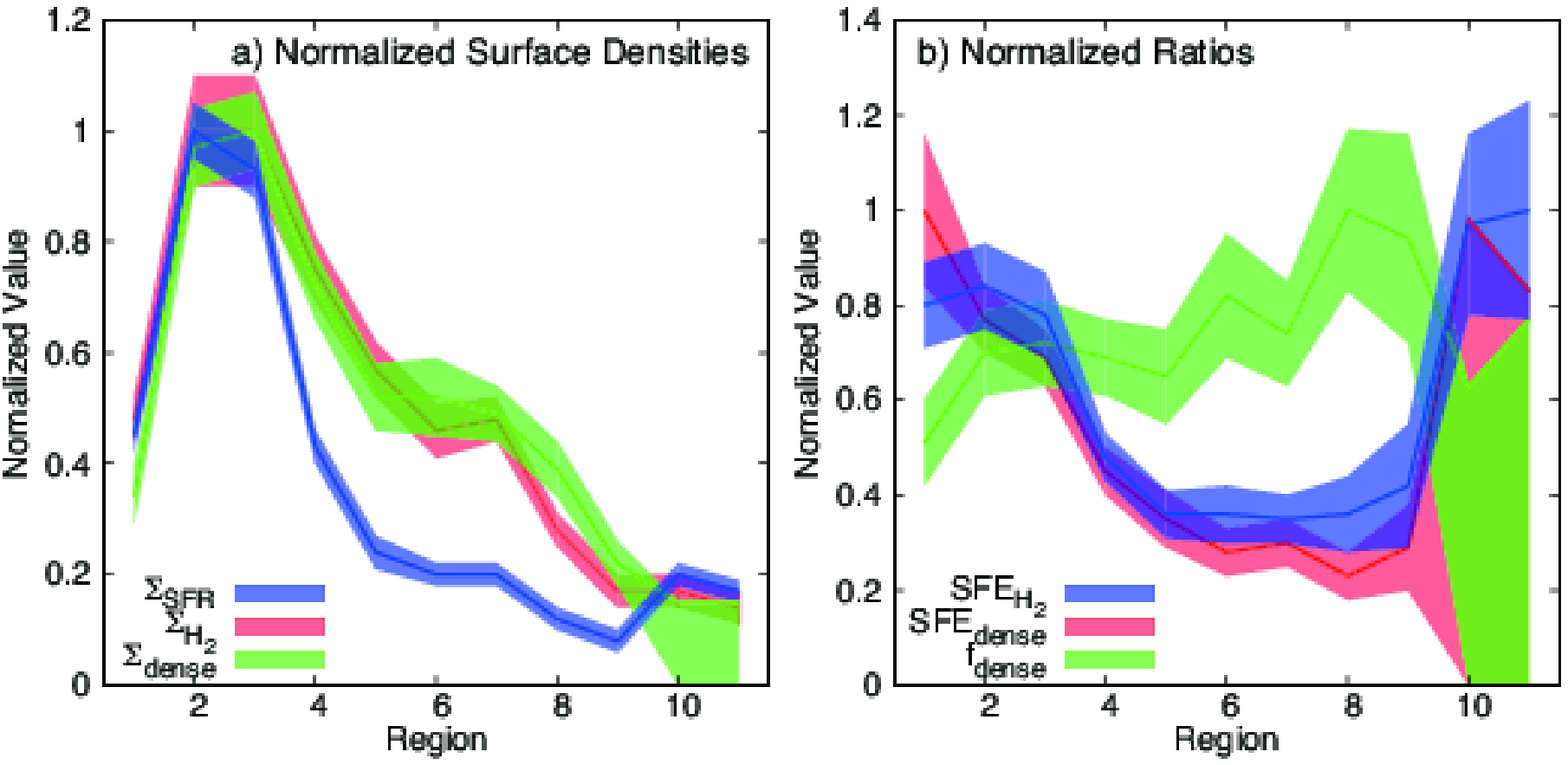}
\caption{(a) Normalized distributions of $\Sigma_{\rm SFR}$ (blue), $\Sigma_{\rm H_2}$ (red), and $\Sigma_{\rm dense}$ (green) along the filament of VV~114.
The absolute values are found in Table~\ref{table_sfe} and Table~2 of \citet{Saito16b}.
(b) Normalized distributions of SFE$_{\rm H_2}$ (blue), SFE$_{\rm dense}$ (red), and $f_{\rm dense}$ (green) along the filament of VV~114.
The absolute values are found in Table~\ref{table_sfe}.
We note that the Overlap region shows higher $f_{\rm dense}$, but lower SFE$_{\rm H_2}$ and SFE$_{\rm dense}$ (see text in detail).
}
\label{fig_21}
\end{center}
\end{figure*}

Considering the observational fact that AGN contribution to the total IR luminosity increases as the IR luminosity increases \citep[e.g.,][]{Ichikawa14}, the presence of AGN can boost higher $J$ HCN and HCO$^+$ fluxes.
However, we exclude the data point which might be dominated by AGN (i.e., Region~1) when deriving the slope, so this is not an applicable explanation for the case of the filament of VV~114.

We note that the $\Sigma_{\rm SFR}$ -- integrated flux plots are equivalent to the FIR luminosity -- dense gas luminosity relations in the sense that all data points shown here can be assumed to have a similar distance, aperture size, and filling factor, and thus, the slope will not change.

\subsection{\texorpdfstring{SFE$_{\rm H_2}$, SFE$_{\rm dense}$, and $f_{\rm dense}$}{SFEH2, SFEdense, and fdense}} \label{SFE}
Here we discuss the star formation activity which takes place in the filament of VV~114 using $\Sigma_{\rm SFR}$, molecular gas mass surface densities ($\Sigma_{\rm H_2}$), and dense gas mass surface densities ($\Sigma_{\rm dense}$).
In addition, we obtain SFE$_{\rm H_2}$ (= $\Sigma_{\rm SFR}$/$\Sigma_{\rm H_2}$), SFE$_{\rm dense}$ (= $\Sigma_{\rm SFR}$/$\Sigma_{\rm dense}$), and dense gas fraction, f$_{\rm dense}$, (= $\Sigma_{\rm dense}$/$\Sigma_{\rm H_2}$).
We use $\Sigma_{\rm H_2}$ as in \citet{Saito16b},
which utilizes the 880~$\mu$m dust continuum emission and the formulation to obtain molecular gas mass from the 880~$\mu$m flux density described in \citet{Scoville16}.
$\Sigma_{\rm dense}$ is calculated by dividing the dense gas mass ($M_{\rm dense}$) by the aperture area.
$M_{\rm dense}$ is calculated using the following equation \citep{Gao&Solomon04b},
\begin{equation}
M_{\rm dense} = \alpha_{\rm HCN(1-0)}L'_{\rm HCN(1-0)}\:M_{\odot}
\end{equation}
where $\alpha_{\rm HCN(1-0)}$ is the HCN~(1--0) luminosity to dense gas mass conversion factor in $M_{\odot}$ (K km s$^{-1}$ pc$^2$)$^{-1}$ and $L'_{\rm HCN(1-0)}$ is the HCN~(1--0) luminosity in K km s$^{-1}$ pc$^2$.
We adopted 10 $M_{\odot}$ (K km s$^{-1}$ pc$^2$)$^{-1}$ for $\alpha_{\rm HCN(1-0)}$ as in \citet{Gao&Solomon04b}.
The derived surface densities are listed in Table~\ref{table_sfe}.

We show surface densities along the filament of VV~114 in Figure~\ref{fig_21}a.
To compare the overall trends between three quantities, we normalized all data points by their peak values.
The normalization of $\Sigma_{\rm SFR}$, $\Sigma_{\rm H_2}$, and $\Sigma_{\rm dense}$ is 56.3~$M_{\odot}$ kpc$^{-2}$ yr$^{-1}$, 1420~$M_{\odot}$ pc$^{-2}$, and 268~$M_{\odot}$ pc$^{-2}$, respectively.
All profiles peak close to the eastern nucleus (Region~2 or 3), and show a decreasing trend toward the western side (Region~11).
This indicates that Region~2 and 3 are sites of ongoing massive, intense star formation (except for the AGN position, region 1).
On the other hand, the Overlap region (Region 5-8) show moderate star formation ($\sim$1/5 of the nuclear star formation) with relatively large amount of diffuse and dense gas ($\sim$1/2 of gas mass or surface density around the eastern nucleus).
This trend becomes clearer when we use the surface density ratios (i.e., SFE$_{\rm H_2}$, SFE$_{\rm dense}$, and $f_{\rm dense}$).
These are listed in Table~\ref{table_sfe}, and shown in Figure~\ref{fig_21}b.
Note that both SFEs correspond to the intercept of the log $L_{\rm IR}$ - log $L'_{\rm line}$ relations with a slope of unity.
SFEs are high around the eastern nucleus (2-3 times higher than those at the Overlap region).
The western side of the filament (Region 11) also shows high SFE$_{\rm H_2}$, which is comparable to that at Region 2 and 3.
Region~11 coincides with strong H$\alpha$ \citep{Zaragoza-Cardiel16} and Pa $\alpha$ \citep{Tateuchi15} peaks, so this may be a relatively unobscured (i.e., not dusty) star-forming region associated with the western galaxy \citep{Grimes06}.
In contrast to the SFEs, $f_{\rm dense}$ peaks at Region~8.
The Overlap region has 1.5-2 times larger $f_{\rm dense}$ than Region 1-3 and 11.
The eastern nucleus is the site of young massive starburst, whereas the Overlap region is dominated by dense gas, and a site of moderate star formation, which has, for instance, $M_{\rm dense}$ = (1.8 $\pm$ 0.2) $\times$ 10$^8$~$M_{\odot}$ (Region~6) of fuel for future star formation.

\subsection{Turbulence-regulated Star Formation} \label{SFmodel}
In order to investigate what physical mechanism governs the observable quantities of star formation at the filament of VV~114, we employ a turbulence-regulated star formation model 
(\citealt{Krumholz&McKee05}; see also \citealt{Usero15} for an application to observing data).
A simple formulation of this model is useful to parameterize the observed quantities (i.e., SFE$_{\rm H_2}$ and $f_{\rm dense}$) with molecular cloud properties.
The equations are,
\begin{eqnarray}
{\rm SFE_{H_2}} &=& \epsilon_{\rm SF}\frac{({\mathcal M}/100)^{-0.32}}{\tau_{\rm ff}(\bar{n})},\\
f_{\rm dense} &=& \frac{1}{2}\left[1+ {\rm erf}\left( \frac{\sigma_{\rm PDF}^2-2\log{(n_{\rm dense}/\bar{n})}}{2^{3/2}\sigma_{\rm PDF}}\right)\right],\\
\sigma_{\rm PDF}^2 &\approx& \log{\left(1+\frac{3{\mathcal M}^2}{4}\right)},
\end{eqnarray}
where $\epsilon_{\rm SF}$ is the efficiency depending on the virial state of the clouds, ${\mathcal M}$ is the Mach number, $\tau_{\rm ff}(\bar{n})$ is the free-fall timescale ($\sqrt{3\pi/(32\bar{n}G)}$) for a given average volume density ($\bar{n}$), and $\sigma_{\rm PDF}$ is the width of the lognormal probability distribution function (PDF) of gas density within a cloud.
This model explains the star formation in a virialized molecular cloud using two parameters, ${\mathcal M}$ and $\bar{n}$: (1) When turbulence in a cloud is high (i.e., high ${\mathcal M}$), it inhibits star formation leading to low SFE$_{\rm H_2}$, although it broadens the density PDF (i.e., increasing the fraction of HCN-emitting dense gas = high $f_{\rm dense}$).
(2) When the average gas density ($\bar{n}$) increases, it shortens $\tau_{\rm ff}(\bar{n})$ (i.e., SFE$_{\rm H_2}$ increases), and naturally results in higher $f_{\rm dense}$.
Throughout this Paper, we assume $n_{\rm dense}$ = 2.8 $\times$ 10$^5$~cm$^{-3}$ and $\epsilon_{\rm SF}$ = 0.28\%, which are the same values as those used in \citet{Usero15}.

\begin{figure}
\begin{center}
\includegraphics[width=8cm]{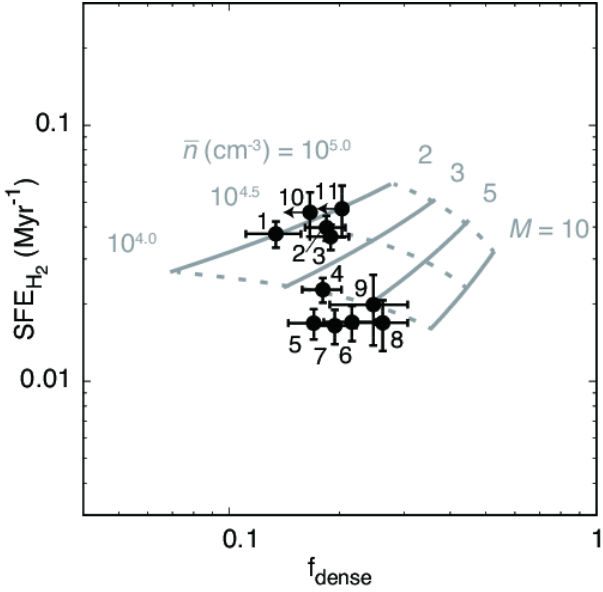}
\caption{SFE$_{\rm H_2}$ plots against $f_{\rm dense}$.  Aperture IDs are shown for each data point.  Grids of the turbulence-regulated star formation model are shown in grey curves \citep{Krumholz&McKee05,Usero15}.  Model curves with constant Mach number (${\mathcal M}$) or constant average density ($\bar{n}$) are shown in solid or dashed curves, respectively.
}
\label{fig_23}
\end{center}
\end{figure}

The observed log SFE$_{\rm H_2}$ against log $f_{\rm dense}$ is plotted in Figure~\ref{fig_23}.
As already seen in Figure~\ref{fig_21}, data points at the Overlap region shows systematically high $f_{\rm dense}$ and low SFE$_{\rm H_2}$.
We overlaid the turbulence-regulated star formation model (${\mathcal M}$-$\bar{n}$ grid) on this Figure.
The grid clearly characterizes the Overlap region as a turbulent lower density ISM (${\mathcal M}$ $\sim$ 5 and $\bar{n}$ $\sim$ 10$^{4.0}$~cm$^{-3}$) relative to the eastern nucleus (${\mathcal M}$ $\sim$ 2 and $\bar{n}$ $\sim$ 10$^{4.5}$~cm$^{-3}$).
The absolute values are uncertain because of the assumed conversion factors to derive the surface densities and the assigned SFR tracer.
For example, when we use 8.4~GHz continuum emission as the SFR tracer (instead of the 110~GHz continuum), SFE$_{\rm H_2}$ decreases by a factor of $\sim$2, hence ${\mathcal M}$ ($\bar{n}$) tends to increase (decrease).
The effect of uncertainties of the conversion factors on SFEs will be discussed later.
However, since the general trend does not change regardless of the employed SFR tracer or adopted conversion factor, the relative differences between the eastern nucleus and the Overlap region also do not change.

Large-scale shocks driven by the interaction between the progenitor's galaxies where proposed by \citet{Saito16b}, which causes an enhancement of methanol abundance at the Overlap region.
We now suggest that the starburst is induced by this shock.
Molecular clouds at the Overlap region of VV~114 are compressed by violent merger-induced shocks, and then the shock-induced intracloud turbulence have started to dissipate, which in part will form new stars.
This view is consistent with the scenario of widespread star formation at the early-to-mid stages of major mergers predicted by numerical simulations \citep[e.g.,][]{Saitoh09}.
The eastern nucleus already shows intense star formation activity, which might be due to efficient gas inflow towards the nucleus at the early-stage of a merger \citep{Iono04b}.

Another model, the so-called density threshold model \citep[e.g.,][]{Gao&Solomon04a,Gao&Solomon04b}, suggests that the rate at which molecular gas turns into stars depends on the dense gas mass within a molecular cloud \citep[i.e., $\Sigma_{\rm SFR}$ $\propto$ $\Sigma_{\rm dense}$ = $f_{\rm dense}\Sigma_{\rm H_2}$; e.g.,][]{Lada12}, and thus it can simply explain the observed linear correlation between $L_{\rm IR}$ and $L'_{\rm HCN(1-0)}$.
It, in turn, predicts a linear correlation between SFE$_{\rm H_2}$ (= $\Sigma_{\rm SFR}$/$\Sigma_{\rm H_2}$) and $f_{\rm dense}$.
However, as we can see in Figure~\ref{fig_23}, VV~114 data do not show such a positive linear correlation (the correlation coefficient is -0.55), showing that the density threshold model is not appropriate for the VV~114 data.
The density threshold model also predicts constant SFE$_{\rm dense}$ from Galactic star-forming region to extragalactic star-forming region (i.e., $\Sigma_{\rm SFR}$ $\propto$ $\Sigma_{\rm dense}$), although the filament of VV~114 shows a decreasing trend towards the west (Figure~\ref{fig_21}).

We note that this result does not change by adopting different conversion factors to derive $\Sigma_{\rm dense}$ and $\Sigma_{\rm H_2}$.
Considering that $\alpha_{\rm HCN}$ scales similarly as $\alpha_{\rm CO}$, and $\alpha_{\rm CO}$ is a few times lower at the nuclear regions than in other quiescent regions of galaxies \citep{Sandstrom13}, $\alpha_{\rm HCN}$ should not be larger at the eastern nucleus of VV~114 than at the Overlap region.
In order to get a constant SFE$_{\rm dense}$ along the filament of VV~114, we could need more (less) $\Sigma_{\rm dense}$ at the eastern nucleus (Overlap region), resulting in 2-3 times larger $\alpha_{\rm HCN}$ at the eastern nucleus, which is not an applicable value.

Similar to SFE$_{\rm dense}$, the difference of SFE$_{\rm H_2}$ between the nucleus and the Overlap region tends to increase when adopting plausible conversion factors.
We used the formulation described in \citet{Scoville16} to derive $\Sigma_{\rm H_2}$ from 880~$\mu$m continuum flux \citep[see][]{Saito16b}, assuming a constant $T_{\rm dust}$ of 25~K.
However, the intense nuclear starburst at the eastern nucleus may result in warmer gas temperature than that for the Overlap region (see Section~\ref{LTE}).
Since the derived $\Sigma_{\rm H_2}$ decreases with increasing $T_{\rm dust}$, SFE$_{\rm H_2}$ tends to increase (decrease) at the eastern nucleus (Overlap region), which enlarges the observed SFE$_{\rm H_2}$ gradient along the filament.
In summary, the systematic uncertainties of the conversion factors do not change the trend seen in Figure~\ref{fig_23}.

\begin{figure*}
\begin{center}
\includegraphics[width=15cm]{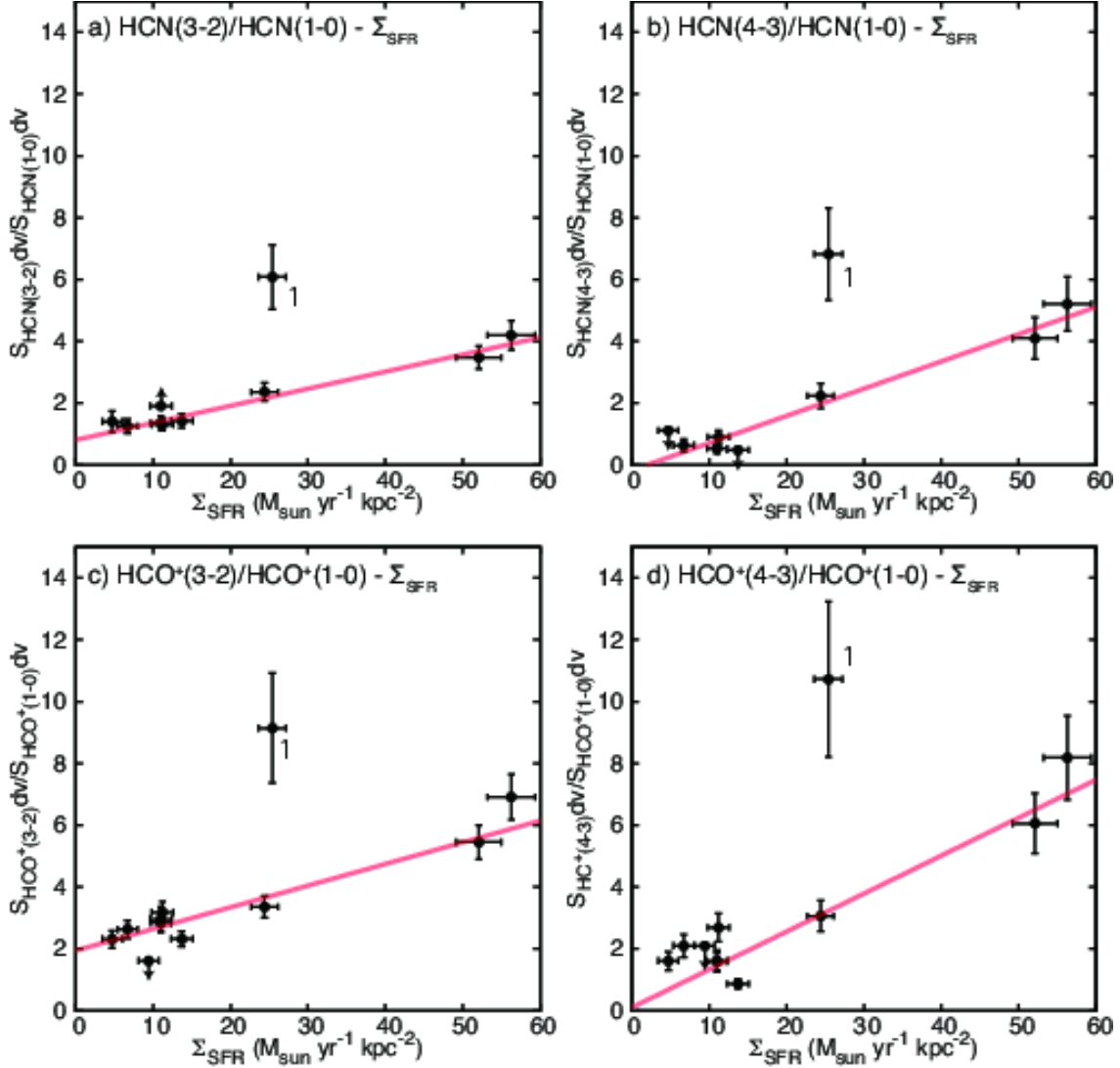}
\caption{Excitation ratio plots against $\Sigma_{\rm SFR}$: (a) HCN~(3--2)/HCN~(1--0) ratio, (b) HCN~(4--3)/HCN~(1--0) ratio, (c) HCO$^+$~(3--2)/HCO$^+$~(1--0) ratio, and (d) HCO$^+$~(4--3)/HCO$^+$~(1--0) ratio.  Data point at the AGN position is denoted as 1.
The best fitted lines except for Region 1 are shown as a red line.
}
\label{fig_15}
\end{center}
\end{figure*}

\subsection{\texorpdfstring{Line Ratios vs. $\Sigma_{\rm SFR}$}{Line Ratios vs. SigmaSFR}} \label{ratio_discuss}
Here we examine the dependence of dense gas excitation ratios on $\Sigma_{\rm SFR}$ using the measured data along the filament of VV~114 (see Table~\ref{table_flux1}), as previously done for CO excitation \citep{Tsai12,Rosenberg15,Kamenetzky16}.
As shown in Figure~\ref{fig_15}, all excitation ratios measured in VV~114 correlate well with $\Sigma_{\rm SFR}$, although Region~1 (i.e., AGN position) clearly deviates from the correlation.
The correlation coefficients are 0.66, 0.70, 0.67, and 0.70, for HCN~(3--2)/HCN~(1--0) (Figure~\ref{fig_15}a), HCN~(4--3)/HCN~(1--0) (Figure~\ref{fig_15}b), HCO$^+$~(3--2)/HCO$^+$~(1--0) (Figure~\ref{fig_15}c), and HCO$^+$~(4--3)/HCO$^+$~(1--0) (Figure~\ref{fig_15}d), respectively, although the coefficients become 0.97, 0.99, 0.96, and 0.95, respectively when excluding Region~1.
This indicates that the excitation condition at Region~1 is clearly different from that at other regions in the filament, and star formation activity may govern dense gas excitation except for Region~1.
A conceivable explanation of the unusual excitation found in Region~1 is the influence of an AGN.
Region~1 shows the highest excitation ratios in VV~114 but modest $\Sigma_{\rm SFR}$.
The $\Sigma_{\rm SFR}$ at Region~1, derived by using 110~GHz continuum emission, may be an upper limit because of the non-thermal contribution from the putative AGN \citep[see][]{Saito16b}, so Region~1 deviates more and more when excluding the possible AGN contribution.
This is consistent with the results from high-resolution observations of multiple molecular lines toward the nearest type~2 AGN host LIRG NGC~1068 \citep{Garcia-Burillo14,Viti14}.
The circumnuclear disk of NGC~1068, whose ISM is thought to be dominantly affected by the central AGN, has at least a few times higher excitation ratios of CO, HCN, and HCO$^+$ than in the starburst ring with hundred pc scale.

\begin{figure*}
\begin{center}
\includegraphics[width=18cm]{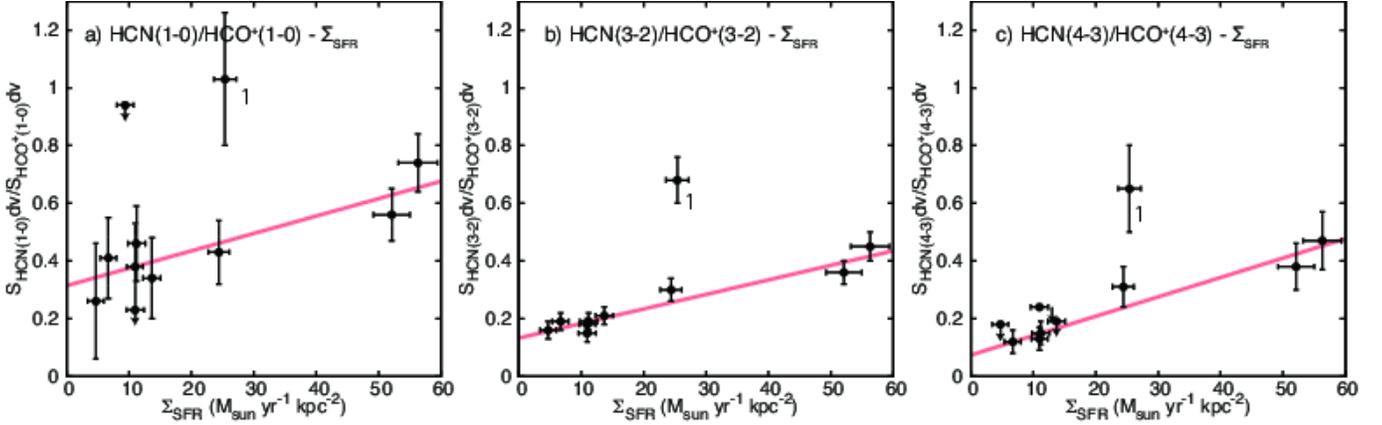}
\caption{HCN/HCO$^+$ ratio plots against $\Sigma_{\rm SFR}$: (a) HCN~(1--0)/HCO$^+$~(1--0) ratio, (b) HCN~(3--2)/HCO$^+$~(3--2) ratio, and (c) HCN~(4--3)/HCO$^+$~(4--3) ratio.  Data point at the AGN position is denoted as 1.
The best fitted lines except for Region 1 are shown as a red line.
}
\label{fig_16}
\end{center}
\end{figure*}

The HCN/HCO$^+$ line ratios show a similar trend against $\Sigma_{\rm SFR}$.
The correlation coefficients are 0.46, 0.57, and 0.62, for HCN~(1--0)/HCO$^+$~(1--0) (Figure~\ref{fig_16}a), HCN~(3--2)/HCO$^+$~(3--2) (Figure~\ref{fig_16}b), and HCN~(4--3)/HCO$^+$~(4--3) (Figure~\ref{fig_16}c), respectively, although the coefficients become 0.85, 0.87, and 0.93, respectively when excluding Region~1.
Since, for a given $J$, HCN has a few times higher critical density than HCO$^+$ \citep{Shirley15}, the strong correlation between HCN/HCO$^+$ ratios and $\Sigma_{\rm SFR}$ is naturally explained by FUV heating from star-forming regions as with the correlation between excitation ratios and $\Sigma_{\rm SFR}$.
On the other hand, Region~1 clearly deviates from trends found in all line ratio plots against $\Sigma_{\rm SFR}$, suggesting the presence of another mechanisms to boost the observed line ratios (i.e., AGN).

We note that Figures~\ref{fig_15} and \ref{fig_16} can be also regarded as luminosity ratio plots against FIR luminosity because all data points have similar distance, aperture size, and filling factor.

\subsection{Radiative Transfer Modeling under LTE} \label{LTE}
Here we derive excitation temperature ($T_{\rm rot}$) and column density ($N_{\rm tot}$) at each aperture position in VV~114 using a single component rotation diagram method, assuming optically-thin gas and LTE conditions.
Recent multiple CO transitions have revealed that nearby IR-bright galaxies \citep[e.g.,][]{Downes&Solomon98,Zhu03,Iono07,Zhu07,Sliwa12,Saito15,Saito17b} show CO~(1--0) emission with moderate optical depth ($\sim$1).
Considering 1-2 orders of magnitude weaker HCN~(1--0) and HCO$^+$~(1--0) emission relative to CO~(1--0) in such IR-bright galaxies \citep{Gao&Solomon04a,Gao&Solomon04b,Privon15}, it is reasonable to assume optically-thin HCN~(1--0) and HCO$^+$~(1--0) lines in VV~114.
In Section~\ref{two-comp}, we examine the appropriateness of the optically-thin and the single component assumptions.

\subsubsection{Rotation Diagram Analysis} \label{RD}
We fitted the observed flux densities by using the following equation \citep[e.g.,][]{Goldsmith&Langer99,Watanabe14,Mangum&Shirley15},
\begin{eqnarray}
W_{\rm \nu} &=& \frac{8\pi^3S\mu_0^2\nu N_{\rm tot}}{3kQ_{\rm rot}}\left\{1-\frac{\exp(h\nu/kT_{\rm rot})-1}{\exp(h\nu/kT_{\rm bg})-1}\right\}\nonumber \\
&& \times \exp\left(-\frac{E_{\rm u}}{kT_{\rm rot}}\right),
\end{eqnarray}
where $W_{\rm \nu}$ is the integrated intensity in units of K km s$^{-1}$ (proportional to the upper state column density, $N_{\rm u}$), $S$  is the line strength, $\mu_0$ is the dipole moment, $\nu$ is the frequency of the transition, $Q_{\rm rot}$ is the rotational partition function, $k$ is the Boltzmann constant, $h$ is the Planck constant, $T_{\rm bg}$ is the cosmic microwave background temperature (= 2.73~K), and $E_{\rm u}$ is the upper state energy.
The molecular line database Splatalogue\footnote{http://www.splatalogue.net/} and the Cologne Database for Molecular Spectroscopy \citep[CDMS\footnote{http://www.astro.uni-koeln.de/cdms/catalog\#partition};][]{Muller01,Muller05} were used to get the transition parameters necessary for the calculation.
For regions without multiple $J$ detections, we used an averaged $T_{\rm rot}$ between Region 3 and 9, where are not contaminated by the putative AGN.

\begin{figure*}
\begin{center}
\includegraphics[width=18cm]{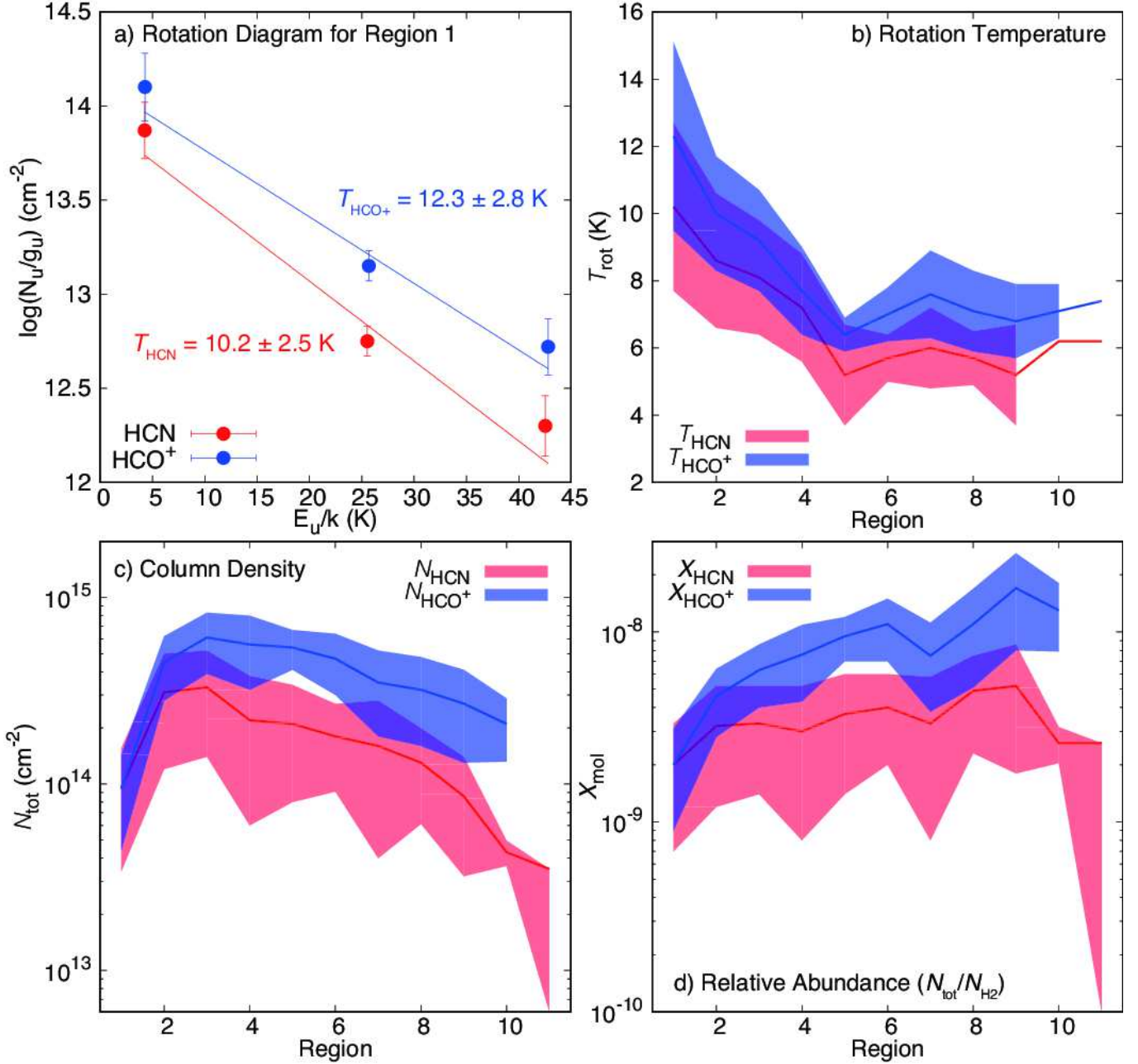}
\caption{(a) Rotation diagrams of HCN and HCO$^+$ for Region~1.  The lines show the best fitted linear functions.  The derived rotation temperature ($T_{\rm rot}$) values are also shown.  (b) $T_{\rm rot}$ obtained for each aperture along the filament of VV~114.  1$\sigma$ error ranges of the fitting are shown as shaded areas.  Values without error indicate the nominal $T_{\rm rot}$.  (c) Column density ($N_{\rm tot}$) along the filament.  (d) Fractional abundance relative to H$_2$ ($X_{\rm mol}$) along the filament.
}
\label{fig_17}
\end{center}
\end{figure*}

\begin{figure}
\begin{center}
\includegraphics[width=8cm]{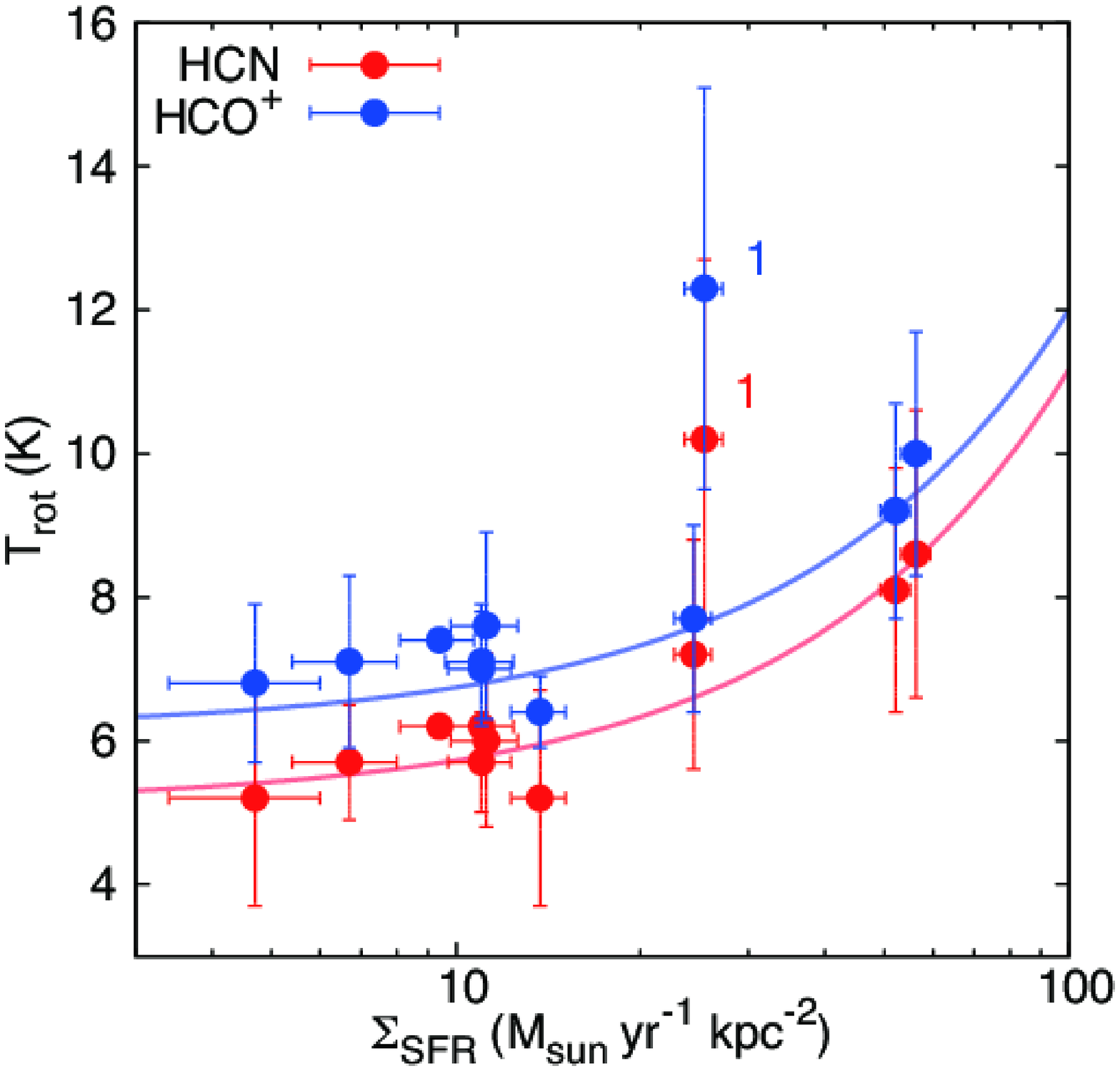}
\caption{$T_{\rm rot}$ plot against $\Sigma_{\rm SFR}$.  Data point at the AGN position is denoted as 1.
The lines show the best fitted linear functions (excluding Region 1).
}
\label{fig_20}
\end{center}
\end{figure}

Figure~\ref{fig_17}a shows the fitting result for Region~1 as an example.
The data appear to be well fitted by a single component, although all regions show that $J$ = 3--2 ($J$ = 1--0 and 4--3) flux is overestimated (underestimated).
Some possibilities to explain this trend will be discussed in Section~\ref{two-comp}.
The derived $T_{\rm rot}$ and $N_{\rm tot}$ for HCN ($T_{\rm HCN}$ and $N_{\rm HCN}$) and HCO$^+$ ($T_{\rm HCO^+}$ and $N_{\rm HCO^+}$) are listed in Table~\ref{table_rotation} and shown in Figures~\ref{fig_17}b and \ref{fig_17}c.
Both $T_{\rm HCN}$ and $T_{\rm HCO^+}$ peak at Region~1 (10.2 $\pm$ 2.5~K and 12.3 $\pm$ 2.8~K, respectively) and decrease towards the western side of the filament.
The derived temperature of 5-8~K at the Overlap region is consistent with the $T_{\rm rot}$ derived from rotational transitions of methanol \citep[$T_{\rm CH_3OH}$ = 6-9~K;][]{Saito16b}.
We compare the derived $T_{\rm HCN}$ and $T_{\rm HCO^+}$ with $\Sigma_{\rm SFR}$ in Figure~\ref{fig_20}.
This figure is more straightforward to interpret than the excitation ratio plots (Figure~\ref{fig_15}) to understand the energetics of dense gas, because some effects controlling the observed line ratios (e.g., optical depth, column density, and fractional abundance) are decomposed.
Both $T_{\rm HCN}$ and $T_{\rm HCO^+}$ correlate with $\Sigma_{\rm SFR}$, except for Region~1.
The correlation coefficients are 0.71 and 0.63 for $T_{\rm HCN}$ and $T_{\rm HCO^+}$, respectively, whereas, without Region~1 data, those are 0.96 and 0.94, respectively.
Those correlations clearly show that HCN and HCO$^+$ molecules are excited by star-forming activities in the filament of VV~114, and Region~1 needs additional efficient heating mechanisms.
This explanation is consistent with CO excitation in (U)LIRGs \citep{Greve14,Kamenetzky16}, indicating that, in general, both diffuse and dense molecular ISM have a similar excitation condition \citep[but see, e.g.,][which show the importance of radiative (IR) pumping for higher-$J$ HCN and HNC]{Sakamoto13,Aalto15b,Imanishi16b}.

\begin{figure}
\begin{center}
\includegraphics[width=8cm]{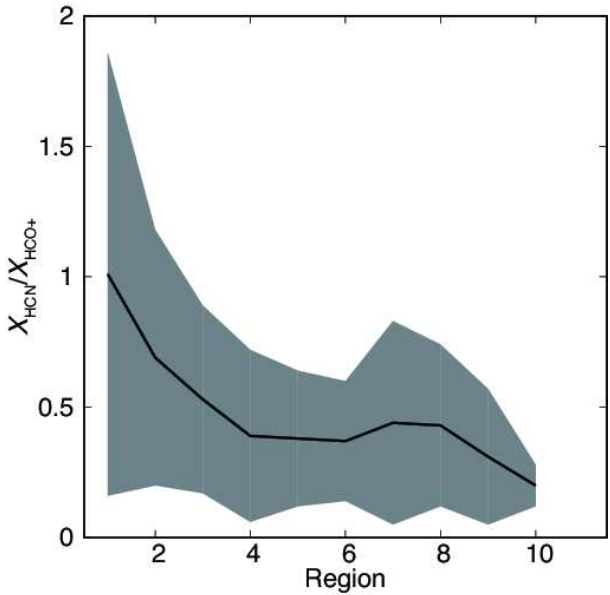}
\caption{$X_{\rm HCN}$/$X_{\rm HCO^+}$ along the filament of VV~114.  The black line shows the best fit value.  The shaded area shows the 1$\sigma$ error range.
}
\label{fig_19}
\end{center}
\end{figure}

The derived dense gas column densities (Figure~\ref{fig_17}c) peak at Region~3 and slightly decrease towards the west, that is, the eastern nucleus contains warmer and more massive dense gas than the Overlap region.
However, when we see the fractional abundances relative to H$_2$ (e.g., $X_{\rm HCN}$ = $N_{\rm HCN}$/$N_{\rm H_2}$), the trend changes.
$X_{\rm HCN}$ shows a remarkably flat distribution along the filament, whereas $X_{\rm HCO^+}$ shows a slightly increasing trend towards the west.
These are the first measurements of spatially-resolved $X_{\rm HCN}$ and $X_{\rm HCO^+}$ for a LIRG.
The average $X_{\rm HCN}$ and $X_{\rm HCO^+}$ are (3.5 $\pm$ 0.7) $\times$ 10$^{-9}$ and (8.9 $\pm$ 1.4) $\times$ 10$^{-9}$, respectively.
Those fractional abundances are similar to other measurements for Galactic star-forming regions and nearby bright galaxies \citep[e.g.,][]{Blake87,Krips08}.
Next, we divide $N_{\rm HCN}$ by $N_{\rm HCO^+}$ to see the spatial distribution of the HCN/HCO$^+$ abundance ratio, $X_{\rm HCN}$/$X_{\rm HCO^+}$, as shown in Figure~\ref{fig_19}.
Although the errors are large, the abundance ratio tends to peak at Region~1.
Also, the HCN~(1--0)/HCO$^+$~(1--0) flux ratio shows a similar trend (Figure~\ref{fig_12}) from 1 (Region~1) to 0.2 (Region~10), indicating that the optically-thin approximation for HCN~(1--0) and HCO$^+$~(1--0) lines is reasonable assumption in this case.

The same analysis for the AGN peak position (E0) shows that $T_{\rm HCN}$ = 9.8 $\pm$ 1.6~K, $N_{\rm HCN}$ = (8.0 $\pm$ 3.8) $\times$ 10$^{13}$~cm$^{-2}$, $T_{\rm HCO^+}$ = 10.9 $\pm$ 2.4~K, $N_{\rm HCO^+}$ = (5.4 $\pm$ 3.3) $\times$ 10$^{13}$~cm$^{-2}$, and $X_{\rm HCN}$/$X_{\rm HCO^+}$ = 1.5 $\pm$ 1.2 (Table~\ref{table_rotation}).
The putative AGN position of VV~114 shows a relatively high $X_{\rm HCN}$/$X_{\rm HCO^+}$, resulting in observed high HCN~(1--0)/HCO$^+$~(1--0) flux ratio.
Higher HCN/HCO$^+$ ratios in higher-$J$ seen at the AGN position is due to the combination effect of the high $X_{\rm HCN}$/$X_{\rm HCO^+}$ and comparably high rotation temperatures between both molecules ($T_{\rm HCN}$  $\simeq$ $T_{\rm HCO^+}$).
Similar rotation temperatures keep the ratio between the upper state column densities of HCN and HCO$^+$ for a given $J$, so the observed HCN/HCO$^+$ ratio SLED seems to be flat close to the AGN position (Figure~\ref{fig_12}).
Other regions along the filament show relatively low $X_{\rm HCN}$/$X_{\rm HCO^+}$ ($\lesssim$ 0.7) and low rotation temperatures ($\lesssim$ 10~K).
Lower HCN temperatures ($T_{\rm HCN}$  $<$ $T_{\rm HCO^+}$) result in the steeper shape of the HCN/HCO$^+$ ratio SLED.

\subsubsection{Two-component ISM?} \label{two-comp}
Here, we discuss the effect of optical depth on the rotation diagram analysis.
According to the LTE radiative transfer calculation described by \citet{Goldsmith&Langer99}, the transition of maximum opacity for a given linear molecule can be written as,
\begin{equation}
J_{\rm max\tau} = 4.6\sqrt{T/B_0},
\end{equation}
where $T$ is the cloud temperature in Kelvin and $B_0$ is the rotation constant in GHz.
For HCN molecule ($B_0$ = 44.31597~GHz) with $T$ = 10-20~K, $J_{\rm max\tau}$ = 2-3.
For the sake of simplicity, we ignore the background radiation term.
In the limit $hB_0/kT$ $<$ 1 (i.e., 0.048$B_0$ $<$ $T$), the ratio of the optical depth of the most optically thick transition ($\tau_{\rm max}$) to that of the $J$ = 1--0 transition ($\tau_{1,0}$) can be expressed as,
\begin{equation}
\frac{\tau_{\rm max}}{\tau_{1,0}} = \frac{T}{0.048eB_0}.
\end{equation}
This indicates that, for HCN molecule with $T$ = 10-20~K, optical depth at $J_{\rm upp}$ $\sim$ 3 is 1.7-3.5 times larger than that at $J_{\rm upp}$ = 1.

On the rotation diagram (for example, Figure~\ref{fig_17}a), data points should be aligned with a single straight line when the observed transitions are optically-thin under LTE.
The expected deviation of the apparent upper state column density at $J_{\rm max\tau}$ from the optically-thin straight line can be estimated using the equation below.
\begin{equation}
N_{\rm u} = \frac{\tau_{\rm max}}{1-e^{-\tau_{\rm max}}}N_{\rm u}^{\rm thin},
\end{equation}
where $N_{\rm u}^{\rm thin}$ is the apparent upper state column density attenuated by line opacity ($\tau_{\rm max}$) and $N_{\rm u}$ is the opacity-corrected upper state column density.
When $\tau$ $\rightarrow$ 0 (i.e., optically-thin limit), $N_{\rm u}^{\rm thin}$ becomes $N_{\rm u}$.
The apparent deviation at $J$ = 3--2 from the straight line through $J$ = 1--0 and 4--3 is $N_{\rm u}/N_{\rm u}^{\rm thin}$ $\sim$ 10$^{13.0}$/10$^{12.75}$ $\sim$ 1.78 (Figure~\ref{fig_17}a), which corresponds to $\tau_{\rm max}$ $\sim$ 1.3.
This $\tau_{\rm max}$ is a lower limit because here we assumed optically-thin $J$ = 1--0 and 4--3 emissions to derive $N_{\rm u}$ = 10$^{13.0}$~cm$^{-2}$.
Using $\tau_{\rm max}/\tau_{1,0}$ = 1.7-3.5, $\tau_{1,0}$ is $\gtrsim$ 0.4-0.8.
Although this is a rough estimate, it is consistent with $\tau_{1,0}$ = 0.3-1.4 derived by the non-LTE modeling for HCN and HCO$^+$ transitions \citep{Krips08}, indicating that the moderate optical depth at $J$ = 3--2 ($\tau_{\rm max}$ $\gtrsim$ 1.3) may drive the apparent two-component excitation seen in the rotation diagram.
This discussion can also be applied to HCO$^+$, which has similar excitation parameters to HCN.
We conclude that the apparent two-component excitation condition seen in the rotation diagrams of HCN and HCO$^+$ can be explained by finite optical depths, which peak at $J_{\rm upp}$ $\sim$ 3.
For much higher-$J$ HCN ($J_{\rm upp}$ $\sim$ 15), radiative pumping rather than collision becomes more important to explain the level population, resulting in a true two-component ISM \citep[e.g.,][]{Rangwala11}.
Observations of other rotational transitions are essential to investigate the appropriateness for other assumptions (e.g., non-LTE excitation).

\section{CONCLUSION} \label{conclusion}
We present new high-resolution (0\farcs2--1\farcs5) images of multiple HCN and HCO$^+$ transitions ($J$ = 1--0, 3--2, and 4--3), HNC~$J$ = 1--0, CS~$J$ = 7--6, and continuum emissions toward the nearby luminous infrared galaxy VV~114 with ALMA bands 3, 6, and 7.
We map the spatial distribution of the three HCN and HCO$^+$ transitions for a LIRG for the first time.
We summarize the main results of this Paper as follows:
\begin{enumerate}
\item All line and continuum images, except for HCN~(4--3) and CS~(7--6) lines (probably because of their high critical densities and low flux densities), show an extended clumpy filament ($\sim$6~kpc length) across the progenitor's galaxy nuclei of VV~114, which has been previously mapped through many molecular and ionized gas tracers.
With our high-resolution Band~7 continuum images ($\sim$80~pc resolution), we found that the filament has a narrow width ($\lesssim$ 200~pc).
Such filamentary structure is predicted by numerical simulations of major mergers.
\item Both excitation ratios (i.e., HCN~($J_{\rm u}$--$J_{\rm l}$)/HCN~(1--0)) and HCN/HCO$^+$ ratios (e.g., HCN~($J_{\rm u}$--$J_{\rm l}$)/HCO$^+$~($J_{\rm u}$--$J_{\rm l}$)) show higher values at the eastern nucleus, relative to the Overlap region between the nuclei of the two galaxies.
We plot spatially-resolved HCN and HCO$^+$ SLEDs for the first time, which clearly shows that the peaks are located at $J_{\rm upp}$ $\lesssim$ 3 for the Overlap region whereas $J_{\rm upp}$ $\gtrsim$ 3 for the eastern nucleus.
The values in the HCO$^+$ SLED are generally larger than those in the HCN SLED.
The HCN/HCO$^+$ ratio SLED shows a remarkably high, flat shape at the AGN position, while a decreasing trend with increasing $J_{\rm upp}$ is seen in other regions.
\item Integrated fluxes - $\Sigma_{\rm SFR}$ and flux ratios - $\Sigma_{\rm SFR}$ plots of different 3$\arcsec$ apertures along the filament show positive trends when excluding the putative AGN, indicating that star formation activity is closely related to surface density and excitation of dense gas.
For both HCN and HCO$^+$, we found that the slope of the log~$L'_{\rm dense}$ -- log~$L_{\rm IR}$ plots decreases as $J$ increases.
This can be explained by subthermalized gas excitation at the lower $\Sigma_{\rm SFR}$ ($L_{\rm IR}$) regime and/or the presence of powerful heating source(s) at the higher $\Sigma_{\rm SFR}$ regime.
\item Furthermore, we estimate the surface densities of total H$_2$ mass and dense gas mass, and then SFE$_{\rm H_2}$ (= $\Sigma_{\rm SFR}$/$\Sigma_{\rm H_2}$), SFE$_{\rm dense}$ (= $\Sigma_{\rm SFR}$/$\Sigma_{\rm dense}$), and $f_{\rm dense}$ (= $\Sigma_{\rm dense}$/$\Sigma_{\rm H_2}$), assuming constant conversion factors.
Although both SFEs peak at the eastern nucleus and are lower at the Overlap region, $f_{\rm dense}$ shows a different trend and peaks at the Overlap region.
This trend becomes clearer when we adopt appropriate conversion factors.
A simple so-called density threshold star formation model cannot reproduce these trends.
\item We examine a turbulence-regulated star formation model, and find that the Overlap region has more diffuse and turbulent dense gas properties relative to that of the eastern nucleus.
Combined with the fact that methanol abundance in gas-phase at the Overlap region is elevated due to large-scale shocks, we suggest that collision between the progenitors of VV~114 has produced the filamentary structure and violent gas inflow toward the eastern nucleus is triggering both starburst and AGN activities.
Merger-induced shocks, which can enhance the intracloud turbulence, dominates at the Overlap region, but dissipates to form new massive star-forming regions.
\item Radiative transfer analysis under LTE and optically-thin assumptions revealed that the rotation temperatures and column densities are higher at the eastern nucleus.
Except for the AGN position, the rotation temperatures are clearly correlated to $\Sigma_{\rm SFR}$, indicating that dense gas excitation in the filament is dominantly governed by star formation activity.
The fractional abundance of HCN relative to H$_2$ is remarkably flat along the filament, showing that the HCN abundance might be insensitive to environmental effects in VV~114 (i.e., AGN, starbursts, and shocked Overlap region).
The average fractional abundances of HCN and HCO$^+$ are (3.5 $\pm$ 0.7) $\times$ 10$^{-9}$ and (8.9 $\pm$ 1.4) $\times$ 10$^{-9}$, respectively.
\item The observed high HCN/HCO$^+$ flux ratio at the eastern nucleus is likely due to a high HCN/HCO$^+$ abundance ratio, and the high excitation ratio there is potentially due to efficient heating by AGN activity.
\end{enumerate}

\acknowledgements

The authors thank an anonymous referee for cements that improved the contents of this paper.
TS and other authors thank ALMA staff for their kind support.
TS acknowledges funding from the European Research Council (ERC) under the European Union’s Horizon 2020 research and innovation programme (grant agreement No. 694343).
This paper makes use of the following ALMA data: ADS/JAO.ALMA\#2011.0.00467.S, ADS/JAO.ALMA\#2013.1.00740.S, ADS/JAO.ALMA\#2013.1.01057.S, and ADS/JAO.ALMA\#2015.1.00973.S.
ALMA is a partnership of ESO (representing its member states), NSF (USA) and NINS (Japan), together with NRC (Canada), NSC and ASIAA (Taiwan), and KASI (Republic of Korea), in cooperation with the Republic of Chile.
The Joint ALMA Observatory is operated by ESO, AUI/NRAO and NAOJ.
This research has made use of the NASA/IPAC Extragalactic Database (NED) which is operated by the Jet Propulsion Laboratory, California Institute of Technology, under contract with the National Aeronautics and Space Administration.


\begin{deluxetable}{lccccccccccc}
\tabletypesize{\scriptsize}
\tablecaption{Line and imaging properties\label{table_data}}
\tablewidth{0pt}
\tablehead{
Line  &$J$ & $\nu_{\rm{rest}}$ & $E_{\rm{u}}/k$ &MRS &$uv$-weight &Beam size &$\Delta V$ &rms & $S_{\rm{line}}\Delta v$ &Recovered flux & Ref. \\
 & & (GHz) & (K) &(\arcsec) &  & (\arcsec) & (km s$^{-1}$) & (mJy b$^{-1}$) & (Jy km s$^{-1}$) & (\%) & \\
(1) &(2) &(3) &(4) &(5) &(6) &(7) &(8) &(9) &(10) &(11) &(12) 
}
\startdata
HCN &1--0 &\phantom{0}88.63160 &\phantom{0}4.25 &22 &briggs &1.45 $\times$ 1.06 &20 &0.4 &\phantom{0}3.53 $\pm$ 0.24 &40 $\pm$ \phantom{0}5 &1 \\
 &3--2 & 265.88643 & 25.52 & \phantom{0}9 & natural & 0.60 $\times$ 0.47 & 30 & 0.4 & \phantom{0}8.36 $\pm$ 0.73 & \nodata & \nodata \\
 &4--3 &354.50548 &42.53 &\phantom{0}7 &natural &0.19 $\times$ 0.15 &30 &0.3 &\phantom{0}5.99 $\pm$ 0.93 & $>$ 18 &2 \\
HCO$^+$ &1--0 &\phantom{0}89.18852 &\phantom{0}4.28 &22 &briggs &1.45 $\times$ 1.04 &20 &0.5 &12.18 $\pm$ 0.69 &52 $\pm$ \phantom{0}4 &1 \\
 &3--2 & 267.55763 & 25.68 & \phantom{0}9 & natural & 0.60 $\times$ 0.47 & 30 & 0.4 & 29.83 $\pm$ 2.44 & \nodata & \nodata \\
 &4--3 &356.73422 &42.80 &\phantom{0}7 &natural &0.19 $\times$ 0.15 &30 &0.4 &18.52 $\pm$ 2.85 & 46 $\pm$ 14 &2 \\
HNC &1--0 &\phantom{0}90.66357 &\phantom{0}4.35 &22 &natural &1.43 $\times$ 1.29 &50 &0.6 & \phantom{0}1.03 $\pm$ 0.20 &27 $\pm$ \phantom{0}7 &1 \\
CS &7--6 &342.88285 &65.83 &\phantom{0}7 &natural &0.18 $\times$ 0.15 &50 &0.3 &\phantom{0}0.41 $\pm$ 0.10 &$>$ 1 & 2
\enddata
\tablecomments{
Column 1: Observed molecule.
Column 2: Observed rotational transition.
Column 3: Rest frequency.
Column 4: Upper state energy.
Column 5: Maximum recoverable scale of the image cube.
Column 6: Visibility ($uv$-plane) weighting for imaging.  ``briggs" means Briggs weighting with robust = 0.5.
Column 7: Beam size of the image cube.
Column 8: Velocity resolution of the image cube.
Column 9: Noise rms in the data which have velocity resolution of $\Delta V$.
Column 10: Total integrated intensity.  We consider the statistical error and the systematic error of the flux calibration in this column.
Column 11: The ALMA interferometric flux divided by the single-dish flux.
Column 12: Reference of the single-dish flux.  (1) \citet{Privon15}; (2) \citet{Zhang14}
}
\end{deluxetable}

\begin{deluxetable}{lccccccccccc}
\tabletypesize{\scriptsize}
\tablecaption{Integrated flux densities inside the 3\farcs0 apertures along the filament of VV~114\label{table_flux1}}
\tablewidth{0pt}
\tablehead{
ID\tablenotemark{a} & $S_{\rm HCN(1-0)}dv$ & $S_{\rm HCN(3-2)}dv$ & $S_{\rm HCN(4-3)}dv$ & $S_{\rm HCO^+(1-0)}dv$ & $S_{\rm HCO^+(3-2)}dv$ & $S_{\rm HCO^+(4-3)}dv$ \\
 & (Jy km s$^{-1}$) & (Jy km s$^{-1}$) & (Jy km s$^{-1}$) & (Jy km s$^{-1}$) & (Jy km s$^{-1}$) & (Jy km s$^{-1}$) 
}
\startdata
1 & 0.38  $\pm$ 0.06  & 2.28  $\pm$ 0.19  & 2.56  $\pm$ 0.41  & 0.36  $\pm$ 0.06  & \phantom{0}3.33  $\pm$ 0.27  & \phantom{0}5.26  $\pm$ 0.80 \\
2 & 1.08  $\pm$ 0.08  & 4.53  $\pm$ 0.37  & 5.62  $\pm$ 0.85  & 1.46  $\pm$ 0.10  & 10.12  $\pm$ 0.81  & 16.08  $\pm$ 2.42 \\
3 & 1.11  $\pm$ 0.08  & 3.87  $\pm$ 0.31  & 4.57  $\pm$ 0.70  & 1.99  $\pm$ 0.12  & 10.85  $\pm$ 0.87  & 16.22  $\pm$ 2.44 \\
4 & 0.80  $\pm$ 0.07  & 1.88  $\pm$ 0.16  & 1.78  $\pm$ 0.28  & 1.85  $\pm$ 0.12  & \phantom{0}6.23  $\pm$ 0.50  & \phantom{0}7.62  $\pm$ 1.15 \\
5 & 0.58  $\pm$ 0.07  & 0.83  $\pm$ 0.09  & $<$ 0.28 & 1.68  $\pm$ 0.12  & \phantom{0}4.25  $\pm$ 0.35  & \phantom{0}1.96  $\pm$ 0.32 \\
6 & 0.58  $\pm$ 0.07  & 0.79  $\pm$ 0.08  & 0.31  $\pm$ 0.08  & 1.52  $\pm$ 0.11  & \phantom{0}4.32  $\pm$ 0.35  & \phantom{0}3.24  $\pm$ 0.50 \\
7 & 0.55  $\pm$ 0.06  & 0.72  $\pm$ 0.07  & 0.49  $\pm$ 0.10  & 1.19  $\pm$ 0.09  & \phantom{0}3.79  $\pm$ 0.31  & \phantom{0}4.31  $\pm$ 0.66 \\
8 & 0.43  $\pm$ 0.05  & 0.54  $\pm$ 0.07  & 0.27  $\pm$ 0.08  & 1.07  $\pm$ 0.08  & \phantom{0}2.80  $\pm$ 0.23  & \phantom{0}3.01  $\pm$ 0.47 \\
9 & 0.24  $\pm$ 0.04  & 0.34  $\pm$ 0.05  & $<$ 0.27 & 0.93  $\pm$ 0.08  & \phantom{0}2.14  $\pm$ 0.18  & \phantom{0}2.00  $\pm$ 0.32 \\
10 & $<$ 0.17 & 0.32  $\pm$ 0.05  & $<$ 0.28 & 0.71  $\pm$ 0.07  & \phantom{0}2.09  $\pm$ 0.18  & \phantom{0}1.53  $\pm$ 0.26 \\
11 & $<$ 0.17 & $<$ 0.26 & $<$ 0.28 & 0.18  $\pm$ 0.06  & $<$ 0.29 & $<$ 0.38  
\enddata
\tablenotetext{a}{Nomenclature as in \citet{Saito16b}.}
\end{deluxetable}

\begin{deluxetable}{lccccccccccc}
\tabletypesize{\scriptsize}
\tablecaption{Flux densities at ``AGN" position (E0)\tablenotemark{a} and ``starburst" position (E1)\tablenotemark{a}\label{table_flux2}}
\tablewidth{0pt}
\tablehead{
 & Unit & AGN (E0) & starburst (E1)
}
\startdata
$S_{254}$ & mJy & 1.50 $\pm$ 0.12 & \nodata \\
$S_{350}$ & mJy & 1.98 $\pm$ 0.30 & 1.49 $\pm$ 0.23 \\
$S_{\rm HCN(1-0)}dv$ & Jy km s$^{-1}$ & 0.34 $\pm$ 0.04 & \nodata \\
$S_{\rm HCN(3-2)}dv$ & Jy km s$^{-1}$ & 1.79 $\pm$ 0.15 & \nodata \\
$S_{\rm HCN(4-3)}dv$ & Jy km s$^{-1}$ & 1.60 $\pm$ 0.24 & 0.62 $\pm$ 0.10 \\
$S_{\rm HCO^+(1-0)}dv$ & Jy km s$^{-1}$ & 0.31 $\pm$ 0.04 & \nodata \\
$S_{\rm HCO^+(3-2)}dv$ & Jy km s$^{-1}$ & 1.53 $\pm$ 0.13 & \nodata \\
$S_{\rm HCO^+(4-3)}dv$ & Jy km s$^{-1}$ & 1.54 $\pm$ 0.24 & 1.20 $\pm$ 0.18 \\
$S_{\rm HNC(1-0)}dv$ & Jy km s$^{-1}$ & 0.29 $\pm$ 0.07 & \nodata \\
$S_{\rm CS(7-6)}dv$ & Jy km s$^{-1}$ & 0.33 $\pm$ 0.07 & $<$ 0.12 \\
$S_{\rm HCN(\nu_2=1^f, 3-2)}dv$ & Jy km s$^{-1}$ &$<$ 0.26 &\nodata \\
$S_{\rm HCN(\nu_2=1^f, 4-3)}dv$ & Jy km s$^{-1}$ &$<$ 0.26 &$<$ 0.12 
\enddata
\tablenotetext{a}{positions defined by \citet{Iono13}}
\end{deluxetable}

\begin{deluxetable}{lccccccccccc}
\tabletypesize{\scriptsize}
\tablecaption{$\Sigma_{\rm dense}$, $f_{\rm dense}$, SFE$_{\rm H_2}$, and SFE$_{\rm dense}$ \label{table_sfe}}
\tablewidth{0pt}
\tablehead{
ID & $\Sigma_{\rm dense}$ & $f_{\rm dense}$ & SFE$_{\rm H_2}$ & SFE$_{\rm dense}$ \\
 & ($M_{\odot}$ pc$^{-2}$) &  & (Myr$^{-1}$) & (Myr$^{-1}$) \\
(1) &(2) &(3) &(4) &(5)
}
\startdata
1 & \phantom{0}90 $\pm$ 13 & 0.13 $\pm$ 0.02 & 0.038 $\pm$ 0.004 & 0.28 $\pm$ 0.05 \\
2 & 260 $\pm$ 20 & 0.18 $\pm$ 0.02 & 0.040 $\pm$ 0.004 & 0.22 $\pm$ 0.02 \\
3 & 268 $\pm$ 18 & 0.19 $\pm$ 0.02 & 0.037 $\pm$ 0.004 & 0.19 $\pm$ 0.02 \\
4 & 192 $\pm$ 17 & 0.18 $\pm$ 0.02 & 0.023 $\pm$ 0.003 & 0.13 $\pm$ 0.01 \\
5 & 139 $\pm$ 17 & 0.17 $\pm$ 0.03 & 0.017 $\pm$ 0.002 & 0.10 $\pm$ 0.02 \\
6 & 140 $\pm$ 18 & 0.22 $\pm$ 0.04 & 0.017 $\pm$ 0.003 & 0.08 $\pm$ 0.01 \\
7 & 132 $\pm$ 14 & 0.19 $\pm$ 0.03 & 0.016 $\pm$ 0.003 & 0.08 $\pm$ 0.01 \\
8 & 104 $\pm$ 13 & 0.26 $\pm$ 0.04 & 0.017 $\pm$ 0.004 & 0.06 $\pm$ 0.01 \\
9 & \phantom{0}59 $\pm$ 11 & 0.26 $\pm$ 0.06 & 0.020 $\pm$ 0.006 & 0.08 $\pm$ 0.03 \\
10 & $<$ 40 & $<$ 0.17 & 0.046 $\pm$ 0.009 & $<$ 0.28 \\
11 & $<$ 41 & $<$ 0.20 & 0.047 $\pm$ 0.011 & $<$ 0.23
\enddata
\tablenotetext{}{
Column 1: Nomenclature as in \citet{Saito16b}.
Column 2: Dense gas mass surface density derived from the HCN~(1--0) integrated intensity.
Column 3: Dense gas fraction (= $\Sigma_{\rm dense}$/$\Sigma_{\rm H_2}$).  We used $N_{\rm H_2}$ as in \citet{Saito16b}, and converted to $\Sigma_{\rm H_2}$.
Column 4: Star formation efficiency (= $\Sigma_{\rm SFR}$/$\Sigma_{\rm H_2}$).  We used $\Sigma_{\rm SFR}$ as in \citet{Saito16b}.
Column 5: Dense gas star formation efficiency (= $\Sigma_{\rm SFR}$/$\Sigma_{\rm dense}$)
}
\end{deluxetable}

\begin{deluxetable}{lccccccccccc}
\tabletypesize{\scriptsize}
\tablecaption{Results of rotation diagram\label{table_rotation}}
\tablewidth{0pt}
\tablehead{
ID\tablenotemark{a} & $T_{\rm HCN}$ & $N_{\rm HCN}$ & $X_{\rm HCN}$ & $T_{\rm HCO^+}$ & $N_{\rm HCO^+}$ & $X_{\rm HCO^+}$ & $X_{\rm HCN}$/$X_{\rm HCO^+}$\tablenotemark{b} \\
 & (K) & (cm$^{-2}$) & (10$^{-9}$) & (K) & (cm$^{-2}$) & (10$^{-9}$) & 
}
\startdata
1 & 10.2 $\pm$ 2.5 & (1.0 $\pm$ 0.6) $\times$ 10$^{14}$ & 2.0 $\pm$ 1.3 & 12.3 $\pm$ 2.8 & (9.4 $\pm$ 5.0) $\times$ 10$^{13}$ & \phantom{0}2.0 $\pm$ 1.1 & 1.0 $\pm$ 0.9 \\
2 & \phantom{0}8.6 $\pm$ 2.0 & (3.1 $\pm$ 1.9) $\times$ 10$^{14}$ & 3.2 $\pm$ 2.0 & 10.0 $\pm$ 1.7 & (4.5 $\pm$ 1.7) $\times$ 10$^{14}$ & \phantom{0}4.6 $\pm$ 1.8 & 0.7 $\pm$ 0.5 \\
3 & \phantom{0}8.1 $\pm$ 1.7 & (3.3 $\pm$ 1.9) $\times$ 10$^{14}$ & 3.3 $\pm$ 1.9 & \phantom{0}9.2 $\pm$ 1.5 & (6.1 $\pm$ 2.2) $\times$ 10$^{14}$ & \phantom{0}6.3 $\pm$ 2.3 & 0.5 $\pm$ 0.4 \\
4 & \phantom{0}7.2 $\pm$ 1.6 & (2.2 $\pm$ 1.6) $\times$ 10$^{14}$ & 3.0 $\pm$ 2.2 & \phantom{0}7.7 $\pm$ 1.3 & (5.6 $\pm$ 2.4) $\times$ 10$^{14}$ & \phantom{0}7.6 $\pm$ 3.3 & 0.4 $\pm$ 0.3 \\
5 & \phantom{0}5.2 $\pm$ 1.5 & (2.1 $\pm$ 1.3) $\times$ 10$^{14}$ & 3.7 $\pm$ 2.3 & \phantom{0}6.4 $\pm$ 0.5 & (5.4 $\pm$ 1.3) $\times$ 10$^{14}$ & \phantom{0}9.5 $\pm$ 2.5 & 0.4 $\pm$ 0.3 \\
6 & \phantom{0}5.7 $\pm$ 0.7 & (1.8 $\pm$ 0.9) $\times$ 10$^{14}$ & 4.0 $\pm$ 2.0 & \phantom{0}7.0 $\pm$ 0.8 & (4.7 $\pm$ 1.7) $\times$ 10$^{14}$ & 10.5 $\pm$ 4.0 & 0.4 $\pm$ 0.2 \\
7 & \phantom{0}6.0 $\pm$ 1.2 & (1.6 $\pm$ 1.2) $\times$ 10$^{14}$ & 3.3 $\pm$ 2.5 & \phantom{0}7.6 $\pm$ 1.3 & (3.5 $\pm$ 1.7) $\times$ 10$^{14}$ & \phantom{0}7.5 $\pm$ 3.7 & 0.4 $\pm$ 0.4 \\
8 & \phantom{0}5.7 $\pm$ 0.8 & (1.3 $\pm$ 0.7) $\times$ 10$^{14}$ & 4.9 $\pm$ 2.6 & \phantom{0}7.1 $\pm$ 1.2 & (3.2 $\pm$ 1.6) $\times$ 10$^{14}$ & 11.5 $\pm$ 5.9 & 0.4 $\pm$ 0.3 \\
9 & \phantom{0}5.2 $\pm$ 1.5 & (8.6 $\pm$ 5.4) $\times$ 10$^{13}$ & 5.2 $\pm$ 3.4 & \phantom{0}6.8 $\pm$ 1.1 & (2.7 $\pm$ 1.4) $\times$ 10$^{14}$ & 16.5 $\pm$ 9.0 & 0.3 $\pm$ 0.3 \\
10 & 6.2\tablenotemark{c} & (4.3 $\pm$ 0.7) $\times$ 10$^{13}$ & 2.6 $\pm$ 5.6 & \phantom{0}7.1 $\pm$ 0.8 & (2.1 $\pm$ 0.8) $\times$ 10$^{14}$ & \phantom{0}1.3 $\pm$ 5.1 & 0.2 $\pm$ 0.1 \\
11 & 6.2\tablenotemark{c} & $<$3.5 $\times$ 10$^{13}$ & $<$ 2.6 & 7.4\tablenotemark{c} & \nodata\tablenotemark{d} & \nodata\tablenotemark{d} & \nodata \\
E0 & \phantom{0}9.8 $\pm$ 1.6 & (8.0 $\pm$ 3.9) $\times$ 10$^{14}$ & \nodata & 10.9 $\pm$ 2.4 & (5.0 $\pm$ 3.0) $\times$ 10$^{14}$ & \nodata & 1.5 $\pm$ 1.2
\enddata
\tablenotetext{a}{Aperture ID defined by \citet{Saito16b}. E0 is the AGN position at the eastern nucleus defined by \citet{Iono13}.}
\tablenotetext{b}{This is derived by $N_{\rm HCN}$/$N_{\rm HCO^+}$ in practice.}
\tablenotetext{c}{In the case of non-detection or single transition detection, we apply the average $T_{\rm rot}$ between Region 3 and 9 to derive $N_{\rm tot}$.  See text.}
\tablenotetext{d}{We did not estimate $N_{\rm HCO^+}$ and $X_{\rm HCO^+}$ at Region 11 because we only marginally detected the HCO$^+$~(1--0) line.}
\end{deluxetable}

\end{document}